\documentclass[floatfix]{emulateapj}
\usepackage{epsfig,subfigure,color,amsmath,graphicx,multirow,subfigmat,ctable}

\bibpunct[; ]{(}{)}{;}{a}{}{,}

\begin{document}

\title{Maximum likelihood analysis of systematic errors in interferometric observations of the cosmic microwave background}
\author{Le Zhang$^{1}$,
 Ata Karakci$^{2}$,
 Paul~M.~Sutter$^{3,4,5,6}$,
 Emory~F.~Bunn$^{7}$,
 Andrei Korotkov$^{2}$,
 Peter Timbie$^{1}$,
 Gregory~S.~Tucker$^{2}$,
 and Benjamin~D.~Wandelt$^{4,5,3,8}$\\
{~}\\
$^{1}$Department of Physics, University of Wisconsin, Madison, WI 53706, USA\\
$^{2}$Department of Physics, Brown University, 182 Hope Street, Providence, RI 02912, USA\\
$^{3}$ Department of Physics, 1110 W Green Street, University of Illinois at Urbana-Champaign, Urbana, IL 61801, USA\\
$^{4}$ UPMC Univ Paris 06, UMR 7095, Institut d'Astrophysique de Paris, 98 bis, boulevard Arago, 75014 Paris, France\\
$^{5}$ CNRS, UMR 7095, Institut d'Astrophysique de Paris, 98 bis, boulevard Arago, 75014 Paris, France\\
$^{6}$ Center for Cosmology and Astro-Particle Physics, Ohio State University, Columbus, OH 43210, USA\\
$^{7}$ Physics Department, University of Richmond, Richmond, Virginia 23173, USA\\
$^{8}$ Department of Astronomy, University of Illinois at Urbana-Champaign, Urbana, IL 61801, USA
}
\thanks{Email: lzhang263@wisc.edu}

\begin{abstract}
We investigate the impact of instrumental systematic errors in interferometric measurements of the cosmic microwave background (CMB) temperature and polarization power spectra. We simulate interferometric CMB observations to generate mock visibilities and estimate power spectra using the statistically optimal maximum likelihood technique. We define a quadratic error measure to determine allowable levels of systematic error that do not induce power spectrum errors beyond a given tolerance. As an example, in this study we focus on differential pointing errors. The effects of other systematics can be simulated by this pipeline in a straightforward manner. We find that, in order to accurately recover the underlying $B$-modes for $r=0.01$ at $28<\ell<384$, Gaussian-distributed pointing errors must be controlled to $0.7^\circ$ rms for an interferometer with an antenna configuration similar to QUBIC, in agreement with analytical estimates. Only the statistical uncertainty for $28<\ell<88$ would be changed at $\sim10\%$ level. With the same instrumental configuration, we find the pointing errors would slightly bias the 2-$\sigma$ upper limit of the tensor-to-scalar ratio $r$ by $\sim10\%$.  We also show that the impact of pointing errors on the $TB$ and $EB$ measurements is negligibly small.
\end{abstract}

\keywords{cosmology:observations, cosmic microwave background, instrumentation:interferometers, methods: data analysis, statistical}

\maketitle

\section{Introduction}
Cosmic microwave background (CMB) polarization measurements can significantly improve the estimation of cosmological parameters, breaking the degeneracies between parameters measured using CMB temperature anisotropy data alone. In the standard theory of the CMB, the polarization field can be decomposed uniquely into an electric-type $E$-mode and a magnetic-type $B$-mode~\citep{zal+seljak:1997,kamionkowski97}. The $E$-mode polarization can provide useful information about reionization of the universe~\citep{Hu2003}. The primordial $B$-modes can probe horizon-scale primordial gravitational waves and play a major role in understanding the inflationary epoch~\citep{Hu2002}, while the secondary lensing-induced $B$-mode signals~\citep{zal+seljak:1998} promise to provide a wealth of information about the distribution of matter and the evolution of large scale structure. Measuring the CMB polarization has become one of the major goals of CMB experiments.
 However, the polarized CMB signal is so small that its measurement requires not only very high instrumental sensitivity, but also exquisite control of systematics.

In many traditional imaging experiments, the determination of the Stokes parameters $Q$ and $U$ is based on subtracting intensities measured by two different detectors. Such an experiment is very sensitive to systematic errors~\citep{hhz:2003}. For instance, beam imperfections or beam mismatch will cause leakage from total intensity $I$ into polarization signals $Q$ and $U$. A recent study~\citep{Miller:2008} shows that in order to achieve a reliable $B$-mode detection ($r = 0.01$), allowable levels of beam systematics should not exceed $1\%$ in ellipticity, the sub-percent level in differential beam width and the few- to sub-arc sec level in differential pointing.
 Also, with a finite patch of sky observed by single-dish intruments, it is impossible to perfectly separate the very weak $B$-modes from the much stronger $E$-modes~\citep{Lewis2002,bunn2002a,bunn2002b,bunn2003,bunn2011}. Therefore polarization detection presents a great challenge in imaging experiments. 

Alternatively, interferometers are a more natural choice for measuring the anisotropies of the CMB temperature and polarization. The correlation of the electric fields from two antennas, called a {\em visibility}, measures the Fourier transform of the sky intensity fluctuations modulated by the response of the antennas. In most cases the region of sky covered by the antennas is small enough that one can use the ``flat-sky'' approximation. The expansion of the intensity field into spherical harmonics thus can be approximated by Fourier modes -- the visibility directly relates to the CMB power spectrum. The main reason for building interferometers instead of traditional imaging experiments is that systematic effects are well-controlled in some cases, especially for $B$-mode detection~\citep{bunnsyst}. Since interferometers measure the Stokes parameters directly -- without differencing the signals between separate detectors for measuring polarizations -- mismatched beam shapes and pointing errors do not cause leakage from $I$ into $Q$ and $U$. In addition, in contrast to imaging experiments, interferometers can separate the $E$- and $B$-modes more cleanly because they sample the sky in the Fourier domain~\citep{Park2003,park2004}.      

Interferometers have been used to measure the CMB anisotropies since the 1980s. The first attempt at measuring the CMB with an interferometer was carried out by~\cite{Martin:1980}. Shortly after that, the 27 antenna VLA  was used in searches for CMB fluctuations~\citep{Fomalont:1984,Knoke:1984,Partridge:1988} and the first dedicated interferometer with two-element correlation receiver to CMB research was made by~\cite{timbie/wilkinson:1988}. So far a number of interferometers have been constructed to observe the CMB power spectrum. The CAT telescope was the first interferometer to actually detect structures in the CMB ~\citep{osullivan/etal:1995,scott/etal:1996,baker/etal:1999} and DASI~\citep{kovac/etal:2002} first detected the faint polarized signals in the CMB.  CBI~\citep{pearson/etal:2003} and VSA~\citep{Dickinson2004,Grainge2003}  made high-sensitivity observations of the CMB temperature and polarization angular power spectra down to sub-degree scales. 

There are many papers in the literature on the study of how instrumental systematic errors affect CMB angular power spectrum measurements. For imaging measurements, a pioneering study of such effects was performed by \citet{hhz:2003}.  In addition, contamination of the CMB power spectrum by systematic effects has been precisely assessed ~\citep{meng/etal:2011,Miller:2008,Dea:2007,shimon/etal:2008,yadav/su/zal:2010}. An analytic approach for characterizing a variety of systematic errors for interferometers has been performed by~\citet{bunnsyst}.   This approach diagnoses systematic errors in a qualitative way.  Actual experiments will naturally require a realistic simulation to quantitatively study such effects as carefully as possible. In this paper, we present for the first time a simulation pipeline to accurately assess the impact of interferometric systematics on CMB power spectrum measurements. Such systematic errors are evaluated from a full maximum likelihood (ML) analysis of realistic simulated data. The method presented in this paper is able to characterize a wide variety of systematic errors, such as beam shape errors, gain errors and cross-polarization, etc. As an example, we examine a specific configuration of an interferometer to quantify systematic pointing errors on the recovery of the CMB power spectrum. 

The remainder of this paper is organized as follows. In Sec.~\ref{sim}, we briefly summarize the simulation of CMB interferometric observations.  In Sec.~\ref{max}, we describe the ML method for extracting the CMB temperature and polarization power spectra from interferometric data. In Sec.~\ref{res}, we focus on pointing errors. Using a comparison of the recovered CMB power spectra with and without pointing errors, we assess the degree of contamination of such systematics on CMB power spectrum recovery. Finally, a discussion and  summary are given in Sec.~\ref{con}.


\section{Simulations}
\label{sim}

\subsection{Visibilities and Covariance Matrix}
Following the previous papers~\citep{bunnsyst,Hobson2002,Hobson1996,Myers2003,MYERS2006,Park2003,White1999}, here we briefly review CMB interferometric observations. Suppose that we have a set of antennas at position ${\bf r_n}$, $n=1,\dotsc$, and each antenna measures two polarization states (linear polarizations or circular polarizations).  The signal received, ${\bf\epsilon}_{out}$, by the $n$-th antenna in response to incoming radiation fields, ${\bf\epsilon}_{in}$, at frequency, $\nu$, from direction ${\bf \hat{k}}$ is then
\begin{equation} 
{\bf\epsilon}_{out} ({\bf r}_n)=\int d^2\hat{\bf k}\, {\bf G}({\bf \hat{k}})\cdot{\bf \epsilon}_{in}({\bf \hat{k}}) 
e^{i({\bf{k}}\cdot {\bf r}_n -2\pi\nu t)} \,.
\label{eq:dataA} 
\end{equation}

Here ${\bf G}$ is the $2\times2$ matrix-valued antenna pattern which encodes the antennas' response to the sky. ${\bf \epsilon}_{in}$ and ${\bf \epsilon}_{out}$ are two-dimensional complex vectors representing an incoming polarized electric field and an electric field that is output by the antenna, respectively. The two-component vectors ${\bf \epsilon}_{in}$ and ${\bf \epsilon}_{out}$ can be expressed in either a linear polarization basis (X-Y) or a circular-polarization basis (R-L). These two bases are connected by a unitary transformation.

The time-averaged value of the correlation between polarization component $m$ from antenna $j$, and polarization component $n$ from antenna $k$ is referred to as a {\em visibility}:  i.e. $V^{jk}_{mn} =\langle\epsilon^j_{out,m}\epsilon^k_{out,n}\rangle$. In the flat-sky approximation, for both the linear and circular-polarization bases, the visibilities with a $2\times2$ matrix form are related to the Stokes parameter matrix as follows:   
\begin{equation} 
V^{jk} =\int d^2{\bf x}\,{\bf G}^j({\bf x})\cdot {\bf S(x)} \cdot{\bf G}^{\dagger k}({\bf x})
e^{-2\pi i{\bf{u}}_{jk}\cdot {\bf x}} \, ,
\label{eq:vis} 
\end{equation}
where the baseline vector ${\bf{u}}_{jk}$ measures the separation between the two antennas in units of wavelength $({\bf r}_j-{\bf r}_k)/\lambda$. The Stokes matrix, ${\bf S}$, is related to the Stokes parameters $I,Q,U,V$ for linear and circular-polarization bases:
\begin{equation} 
{\bf S}_{lin}= \left( \begin{array}{cc} I+Q & U+iV \\
 U-iV &I-Q \end{array}  \right) \, , 
\end{equation}
\begin{equation} 
{\bf S}_{circ}= \left( \begin{array}{cc} I+V & Q+iU \\
 Q-iU &I-V \end{array}  \right) \, .
\end{equation} 

 Usually two main descriptors for systematics are used~\citep{bunnsyst}: the instrumental Jones matrix, ${\bf J}$, and the antenna pattern, ${\bf G(x)}$. ${\bf J}$ describes systematics that are purely introduced within the instrument, such as gain errors and cross-talk between the two outputs of a given antenna. Instrumental errors are easier to model since they do not depend on the position on the sky, while ${\bf G(x)}$ characterizes systematics occurring from the observation of the sky before instrumental errors are taken into account. We thus use ${\bf G(x)}$ to model beam shape errors, pointing errors and cross polarizations, etc. This separation is a bit arbitrary in that instrumental errors could be absorbed into antenna patterns, but they are a convenient conceptual distinction between systematics happening ``before'' and ``after'' the antenna averages over the beam.  The total effect on the visibilities can be found from the relations:
\begin{equation} 
V^{jk} =\int d^2{\bf x}\, {\bf J}^j\left({\bf G}^j({\bf x})\cdot {\bf S(x)} \cdot{\bf G}^{\dagger k}({\bf x})\right) {\bf J}^{\dagger k}
e^{-2\pi i{\bf{u}}_{jk}\cdot {\bf x}} \, .
\label{eq:jones} 
\end{equation} 

For the $j$-th antenna with gain errors $g^j_1$, $g^j_2$ and cross-talk couplings $\epsilon^j_1$, $\epsilon^j_2$, the Jones matrix reads
\begin{equation} 
{\bf J}^j  = \left( 
\begin{array}{cc}
 1+g^j_1  &  \epsilon^j_1\\
 \epsilon^j_2  &  1+g^j_2
 \end{array} \right)\, .
\label{eq:Jerror} 
\end{equation}

Similar to the Jones matrix, an azimuthally symmetric antenna pattern has the form of
\begin{equation} 
{\bf G}^j(r,\phi)  = \left( 
\begin{array}{cc}
 G^j_0+ \frac{1}{2}G^j_1\cos(2\phi)  & \frac{1}{2} G^j_1\sin(2\phi)\\
 \frac{1}{2} G^j_1\sin(2\phi)  &   G^j_0 -\frac{1}{2}G^j_1\cos(2\phi) 
 \end{array} \right)\, ,
\label{eq:Aerror} 
\end{equation}  
where $(r,\phi)$ are polar coordinates, $G_0$ is the ideal beam shape and $G_1$ leads to two polarization states mixing. The scalar functions $G_0$, $G_1$ depend only on $r$. For an ideal interferometer, each antenna has an identical response to both polarization states while there is no mixing between them. In this case, ${\bf G}$ is equal to a scalar function multiplied by the identity matrix, i.e., ${\bf G(x)} =G({\bf x})\mathbf{1}$.

For both linear and circular polarization experiments, coupling errors ($\epsilon$) are the major sources of systematics affecting the $B$-mode power spectrum due to leakage from $I$ into $Q$ and $U$, while less worrisome gain errors would mix $V_Q$ and $V_U$ with each other. Furthermore, as shown by~\cite{bunnsyst}, the visibility for Stokes $V$ could provide a useful diagnostic to monitor the presence of these systematics.    

For simplicity we assume that instrument errors in the Jones matrix are negligible (taking ${\bf J} = \mathbf{1}$) and each antenna has identical beam patterns for both polarization states and has no cross-polar response (i.e., off-diagonal entries in ${\bf G}$). Then, each visibility measures a simple linear combination of the Stokes parameters. We can extract Stokes visibilities from Eq.~\ref{eq:vis}, yielding  
\begin{equation} 
V_Z^{jk} =\int d^2x\, Z({\bf x})G^j({\bf x})G^{*k}({\bf x})
e^{-2\pi i{\bf{u}}_{jk}\cdot {\bf x}} \, ,
\label{eq:eqz} 
\end{equation}  
for $Z = \{I,Q,U,V\}$. Since the visibility function is the Fourier transform of the Stokes fields on the sky weighted by the antenna response, by using the well-known convolution theorem, we can write the visibility function in Fourier space ($uv$-domain):
\begin{equation} 
V^{jk}_Z({\bf u}) =\int d^2w\, \tilde{Z}({\bf u- w})\tilde{G}^{jk}({\bf w}) \, ,
\label{eq:eqfz} 
\end{equation}
 where $\tilde{Z}({\bf u} )$ and $\tilde{G}^{jk}({\bf u})$ are the Fourier transforms of $Z({\bf x})$ and $G^j({\bf x})G^{*k}({\bf x})$, respectively. Note that each antenna pattern could differ from the others because of systematic beam errors.

In this study, we only focus on differential pointing errors. 
The pipeline presented in this paper can be applied in a straightforward manner to other systematics. Pointing errors occur when not all the antennas point in the same direction. Following~\cite{bunnsyst}, we model the pointing error as a Gaussian-distributed error with dispersion $\delta$. Assuming the $j$-th antenna has the pointing offset $\delta{\bf x}_j$ relative to a desired direction, then according to Eq.~\ref{eq:eqz}, each visibility can be expressed in the form
\begin{eqnarray} 
V_Z^{jk} =&&\int d^2x\,Z({\bf x})G^j({\bf x}+\delta{\bf x}_j )G^{*k}({\bf x}+\delta{\bf x}_k) \nonumber \\
          &&\times e^{-2\pi i{\bf{u}}_{jk}\cdot {\bf x}} \, , 
\label{eq:eqzp} 
\end{eqnarray}  
where $Z= \{I,Q,U\}$ and the beam response $G({\bf x})$ can be approximated by a circular Gaussian of a frequency-dependent dispersion $\sigma$, i.e., $G({\bf x}) = \exp(-|{\bf x}|^2/2\sigma^2(\nu))$. The corresponding Fourier transform is thus given by $\tilde{G}({\bf u}) = 2\pi\sigma^2(\nu)\exp(-2\pi^2|{\bf u}|^2\sigma^2(\nu))$. It is worth noticing that the {\em differential pointing} in this paper specifically refers to pointing offsets in some antennas relative to a desired direction. Our definition naturally includes the relative displacement of the beam centroids of two antennas and the average position of the beam centroids, which are usually respectively referred to as ``differential pointing'' and ``common pointing'' in imaging experiments.

Additionally, for azimuthally asymmetric antennas, the pointing offsets for
the two polarization states could be different. Such "non-identical"
pointing errors are much more worrisome in imaging experiments since the
differential pointing effects couple $T$ to $Q$ and $U$ and so produce a large
bias on the $B$-modes~\citep{Miller:2008}. For interferometers, they are
expected to produce contamination at smaller level comparable to that from
identical pointing errors since the biases induced by the offsets in these
two situations both arise from leakage of $E$ into $B$. For simplification, in
this study we assume the pointing offsets $\delta {\bf x}$  are identical for the two polarization states in an arbitrary antenna but can be different for two
different antennas. We will perform a detailed simulation in a forthcoming
paper to quantitatively assess effect of non-identical pointing errors for
interferometers.

In order to recover the power spectrum based on simulated visibility data, one needs to construct the covariance matrix, which is the fundamental tool for analysis of Gaussian random CMB fields. In principle, all kinds of systematic errors in visibilities can be simulated through Eq.~\ref{eq:jones}, whereas the theoretically predicted covariance matrix does not include any systematic uncertainties. Taking the error-free beam pattern, ${\bf G}$, with identical response between antennas, Eqs.~\ref{eq:eqz} and~\ref{eq:eqfz} can be simplified as follows:
\begin{eqnarray} 
V_Z({\bf u}) &=&\int d^2x\, Z({\bf x})A({\bf x})
e^{-2\pi i{\bf{u}}\cdot {\bf x}} \\
V_Z({\bf u}) &=&\int d^2w\, \tilde{Z}({\bf u- w})\tilde{A}({\bf w}) \, .
\label{eq:eqfsz} 
\end{eqnarray}
Here the intensity beam pattern, $A({\bf x})$, is defined by $|G({\bf x})|^2$ and $\tilde{A}$ is its Fourier transform.

In the flat-sky approximation, the Stokes parameters $Q$ and $U$ can be decomposed into  the $E$- and $B$-modes in Fourier space~\citep{zal+seljak:1997}. Using Eq.~\ref{eq:eqfsz}, the Stokes visibilities $V_I,V_Q$ and $V_U$ then can be expressed in terms of $T$, $E$ and $B$ modes as follows:  
\begin{eqnarray}
V_Q({\bf u})&=&\int d^2w\, [\tilde{E}({\bf w})\cos(2\phi_{\bf w}) - \tilde{B}({\bf w})\sin(2\phi_{\bf w})] \tilde{A}({\bf u- w})   \nonumber\\
V_U({\bf u})&=&\int d^2w\, [\tilde{E}({\bf w})\sin(2\phi_{\bf w}) + \tilde{B}({\bf w})\cos(2\phi_{\bf w})] \tilde{A}({\bf u - w}) \nonumber\\
V_I({\bf u})&=&\int d^2w\, \tilde{T}({\bf w}) \tilde{A}({\bf u- w})\, ,
\label{eq:VEB} 
\end{eqnarray}
where $\tilde{T}$, $\tilde{E}$ and $\tilde{B}$ stand for the CMB temperature field, $T$, and polarization fields, $E$ and $B$, in Fourier space and $\phi_{\bf w}$ is the angle made by the vector ${\bf w }$ with respect to the $x$-axis. In this study we assume the Stokes visibility $V_V$ to be zero.

Given a set of visibility measurements, one can use maximum likelihood analysis to evaluate the CMB power spectra of the temperature and polarization. By defining a vector of data ${\bf V}\equiv (V^1_I,V^1_Q,V^1_U;\dotsb;V^n_I,V^n_Q,V^n_U)$ at each baseline vector ${\bf u}_i$ with $i=1,\dotsc,n$, the corresponding covariance matrices of the CMB visibilities are    
\begin{eqnarray} 
C^{ij}_{ZZ'} &\equiv& \left<V_Z({\bf u}_i)V^*_{Z'}({\bf u}_j)\right> \nonumber \\
&=&\int d^2w\int d^2w' \left<\tilde{Z}({\bf w})\tilde{Z'^*}({\bf w'})\right> \nonumber \\ 
  &&{}\times \tilde{A}({\bf u}_i-{\bf w})\tilde{A}^*({\bf u}_j-{\bf w'})\nonumber \\
  &=& \int d^2w\, \mathcal{S}_{zz'}(|{\bf w} |) \tilde{A}({\bf u}_i-{\bf w})\tilde{A}^*({\bf u}_j-{\bf w}) \, ,
\label{eq:eqcz} 
\end{eqnarray}
where $i,j$ denote visibility data indices and the dependence of the correlation functions $\mathcal{S}_{zz'}$ on the CMB power spectra are listed in Table~\ref{tab:hresult}. In the flat sky approximation, the 2D power spectrum $4\pi^2|{\bf u}|^2S(|{\bf u}|)\simeq \ell(\ell+1)C_\ell|_{\ell=2\pi{ \bf u}}$  for $\ell\gtrsim10$~\citep{White1999}.

\begin{table*}[t]
\caption{Dependence of ensemble-averaged Stokes parameter correlations on the CMB angular power spectra} 
\centering
\renewcommand{\arraystretch}{1.5}
\begin{tabular}{c l}
\hline\hline
ZZ'& $\left<\tilde{Z}({\bf w})\tilde{Z'^*}({\bf w'})\right>= \mathcal{S}_{zz'}(|{\bf w}|)\delta (\bf{w-w'})$ \\
\hline 
$II$  & $\mathcal{S}_{II} = S_{TT}(w)$ \\

$IQ$  & $\mathcal{S}_{IQ} = S_{TE}(w) \cos 2\phi_{\bf w} -S_{TB}\sin 2\phi_{\bf w} $ \\

$IU$  & $\mathcal{S}_{IU} =  S_{TE}(w) \sin 2\phi_{\bf w} +S_{TB}\cos 2\phi_{\bf w}$\\

$QQ$  & $\mathcal{S}_{QQ} = S_{EE}(w) \cos^2 2\phi_{\bf w} + S_{BB}(w) \sin^2 2\phi_{\bf w}- S_{EB}(w) \sin 4\phi_{\bf w}$ \\
  
$QU$  & $\mathcal{S}_{QU} = (S_{EE}(w) - S_{BB}(w))\sin 2\phi_{\bf w} \cos 2\phi_{\bf w}  + S_{EB} (\cos^2 2\phi_{\bf w} -\sin^2 2\phi_{\bf w})$\\ 

$UU$  & $\mathcal{S}_{UU} = S_{EE}(w) \sin^2 2\phi_{\bf w} + S_{BB}(w) \cos^2 2\phi_{\bf w} + 2S_{EB}(w)\sin 2\phi_{\bf w} \cos 2\phi_{\bf w}$ \\
\hline
\end{tabular}
\label{tab:hresult}
\end{table*}

\begin{table*}[t]
\label{tab:windows}
\centering
\caption{Table of integrals used in the calculations of the window functions.}
\renewcommand{\arraystretch}{1.5}
\begin{tabular}{l c}
\hline
\hline 
$\int_0^{2\pi} d\phi_{\bf w}\, \exp(4\pi^2{\bf q}\cdot{\bf w}) $ = $2 \pi  I_0(a)$ \\

$\int_0^{2\pi} d\phi_{\bf w}\, \exp(4\pi^2{\bf q}\cdot{\bf w})\cos 2\phi_{\bf w}$  = $2 \pi  I_2(a) \cos (2 \phi_{\bf q}) $ \\

$\int_0^{2\pi} d\phi_{\bf w}\, \exp(4\pi^2{\bf q}\cdot{\bf w})\sin 2\phi_{\bf w}$  = $2 \pi  I_2(a) \sin (2  \phi_{\bf q}) $ \\

$\int_0^{2\pi} d\phi_{\bf w}\, \exp(4\pi^2{\bf q}\cdot{\bf w})\cos^2 2\phi_{\bf w}$  = $\frac{\pi  \left(a I_0(a) \left(\left(a^2+24\right) \cos (4 \phi_{\bf q})+a^2\right)-8 \left(a^2+6\right) I_1(a) \cos (4\phi_{\bf q})\right)}{a^3}$ \\
  
$\int_0^{2\pi} d\phi_{\bf w}\, \exp(4\pi^2{\bf q}\cdot{\bf w})\sin^2 2\phi_{\bf w}$  = $\frac{\pi  \left(8 \left(a^2+6\right) I_1(a) \cos (4 \phi_{\bf q})+I_0(a) \left(a^3-a \left(a^2+24\right) \cos (4\phi_{\bf q})\right)\right)}{a^3}$\\ 

$\int_0^{2\pi} d\phi_{\bf w}\, \exp(4\pi^2{\bf q}\cdot{\bf w})\sin 2\phi_{\bf w}\cos 2\phi_{\bf w}$ =$\frac{\pi  \left(a \left(a^2+24\right) I_0(a)-8 \left(a^2+6\right) I_1(a)\right) \sin (4 \phi_{\bf q})}{a^3}$ \\
\hline
where we introduce $|{\bf q}| =\sigma({\bf u}_i + {\bf u}_j)$, $a =4\pi^2|{\bf q}||{\bf w}|$ and  
$I_m(a)$ is the modified Bessel function\\ of the first kind and order $m$ for a real argument $a$ (see details in the text).
\end{tabular}
\end{table*}

\subsection{Simulated Observations}
The CMB Stokes fields are believed to be isotropic and Gaussian in the standard  inflationary models~\citep{Guth:1981,kamionkowski97,kamionkowski97b,zal+seljak:1997}.  On a small patch of the sky, the corresponding Fourier components of these fields are complex random variables and the value of the real and imaginary parts of each point ${\bf u}$  in Fourier space are drawn independently from a normal distribution with zero mean and variance~$\propto C_\ell|_{\ell=2\pi |{\bf u}|}$. With cosmological parameters derived from WMAP 7-year results~\citep{Larson2011,Komatsu2011}, we use the public code CAMB~\citep{Lewis2000} to compute the CMB power spectra $C^{TT}_\ell,C^{EE}_\ell,C^{TE}_\ell,C^{BB}_\ell$. The input $B$-mode contains a primordial component with a tensor-to-scalar perturbation ratio, $r$, and a secondary component induced by lensing.  In our simulation we fix $r=0.01$, the goal for many current observations.

Based on these power spectra, we generate Fourier modes and then perform the inverse Fourier transform to obtain real-space Stokes fields $I({\bf x}), Q({\bf x})$ and $U({\bf x})$. From Eq.~\ref{eq:jones}, for a given Jones matrix and beam response ${\bf G}$, the Stokes visibilities are then obtained by performing the Fourier transform again.
 
We assume the instrumental noise at each point of the $uv$-plane is a complex, Gaussian-distributed number which is independent between different baselines~\citep{White1999,Morales:2010}. For an instrument which measures both polarizations with an identical uncertainty in Stokes parameters, we can separately generate the Gaussian noise with identical rms levels for each $I$, $Q$ and $U$ visibility. The correlation function of the noise for baselines $i$ and $j$ is determined by 
 \begin{equation} 
C^{ij}_N = \left(\frac{{\lambda}^2T_{sys} }{\eta_A A_D}\right)^2 \left(\frac{1}{\Delta_\nu t_an_b} \right)\delta_{ij}                     \, ,
\label{eq:CN} 
\end{equation}      
where $T_{sys}$ stands for the system noise temperature, $\lambda$ for the observing wavelength, $A_D$ for the physical area of a antenna, $\eta_A$ for the aperture efficiency, $\Delta_\nu$ for bandwidth, $n_b$ for the number of baselines with the same baseline vector ${\bf u}$ and $t_a$ for the integration time of the baseline.

In order to illustrate the effect of systematic errors on the recovered CMB power spectra and set allowable tolerance levels for those errors, we perform simulations for a specific interferometer design.  We choose an antenna configuration similar to that of the QUBIC instrument ~\citep{battistelli/etal:2011} which is under construction for observations at 150 GHz. In our simulation, the interferometer is a two-dimensional square close-packed array of 400 horn antennas with Gaussian beams of width $5^\circ$ in the intensity beam pattern $A({\bf x})$, corresponding to $\sim7.1^\circ$ in $G({\bf x})$.   The antennas have uniform physical separations of $7.89 \lambda$.  With this configuration, the resolution in the $uv$-plane is about $\sigma_u=1.82$ ($\Delta \ell\simeq11$), and the $uv$ coverage reaches down to $\ell\gtrsim 50-2\Delta \ell = 28$, probing the primordial $B$-mode bump at $\ell\approx 50$.

We also assume that all Stokes visibilities. $I$, $Q$ and $U$, can be measured simultaneously for each antenna pair with an associated rms noise level of $0.015\mu$K per visibility, roughly corresponding to low-noise detectors each with $150\mu$Ks$^{1/2}$ and a total integration time of three years. With this noise level, the simulations show that the averaged overall signal-to-noise ratio (SNR) in Stokes $Q$ and $U$ maps is about $5$. The high SNR ensures an accurate recovery of the $B$-mode power spectrum and allows us to see systematic effects clearly.

\begin{figure*}[t]
\centering
\mbox{
\subfigure[$I({\bf x})$]{
   \includegraphics[scale =0.35] {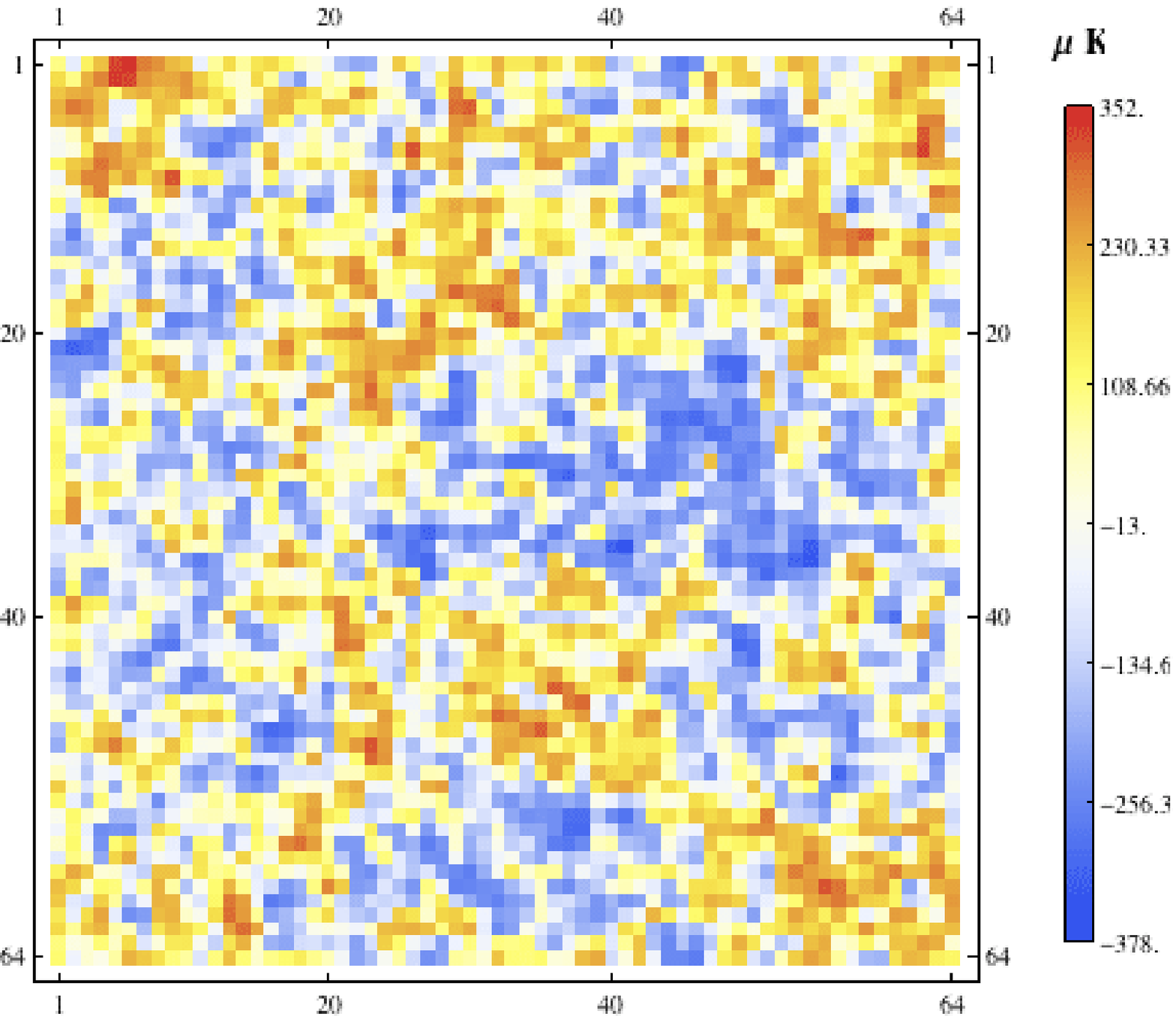} 
   \label{fig:realI} }

 \subfigure[$Q({\bf x})$]{
   \includegraphics[scale =0.35] {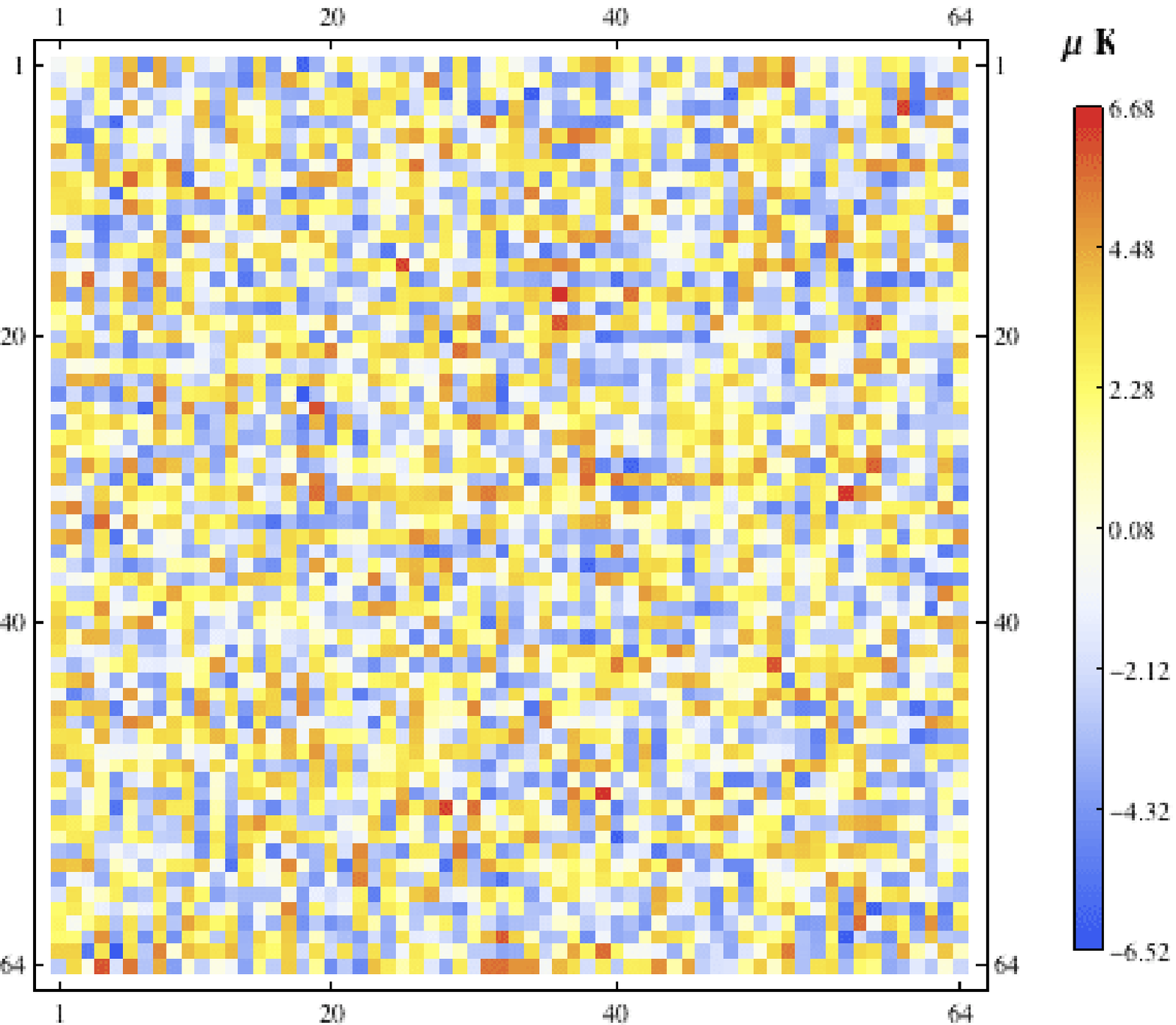}
   \label{fig:realQ} }  

 \subfigure[$U({\bf x})$]{
   \includegraphics[scale =0.35] {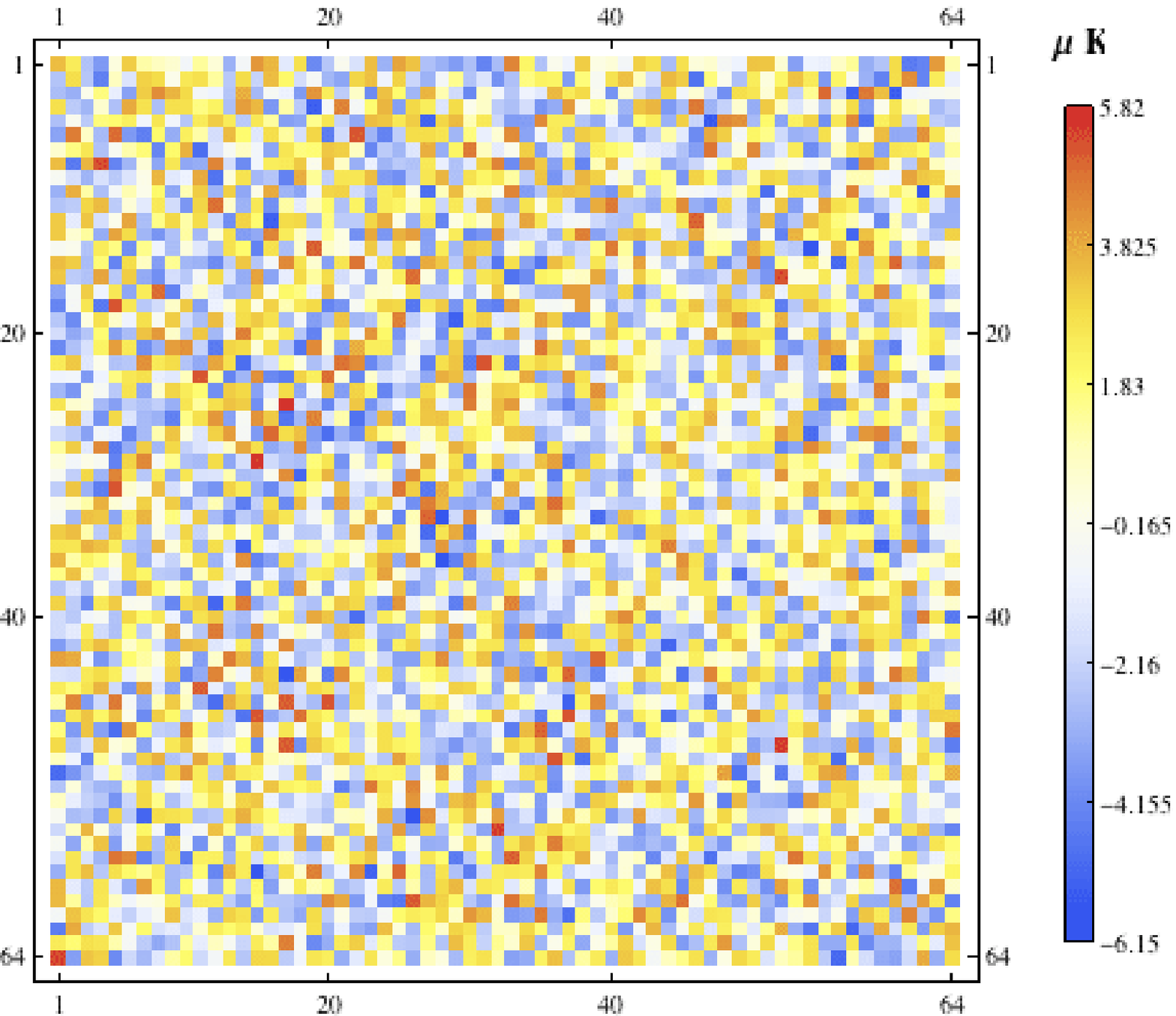}
   \label{fig:realU} }
}

\mbox{
\subfigure[$V_I({\bf u})$]{
   \includegraphics[scale =0.35] {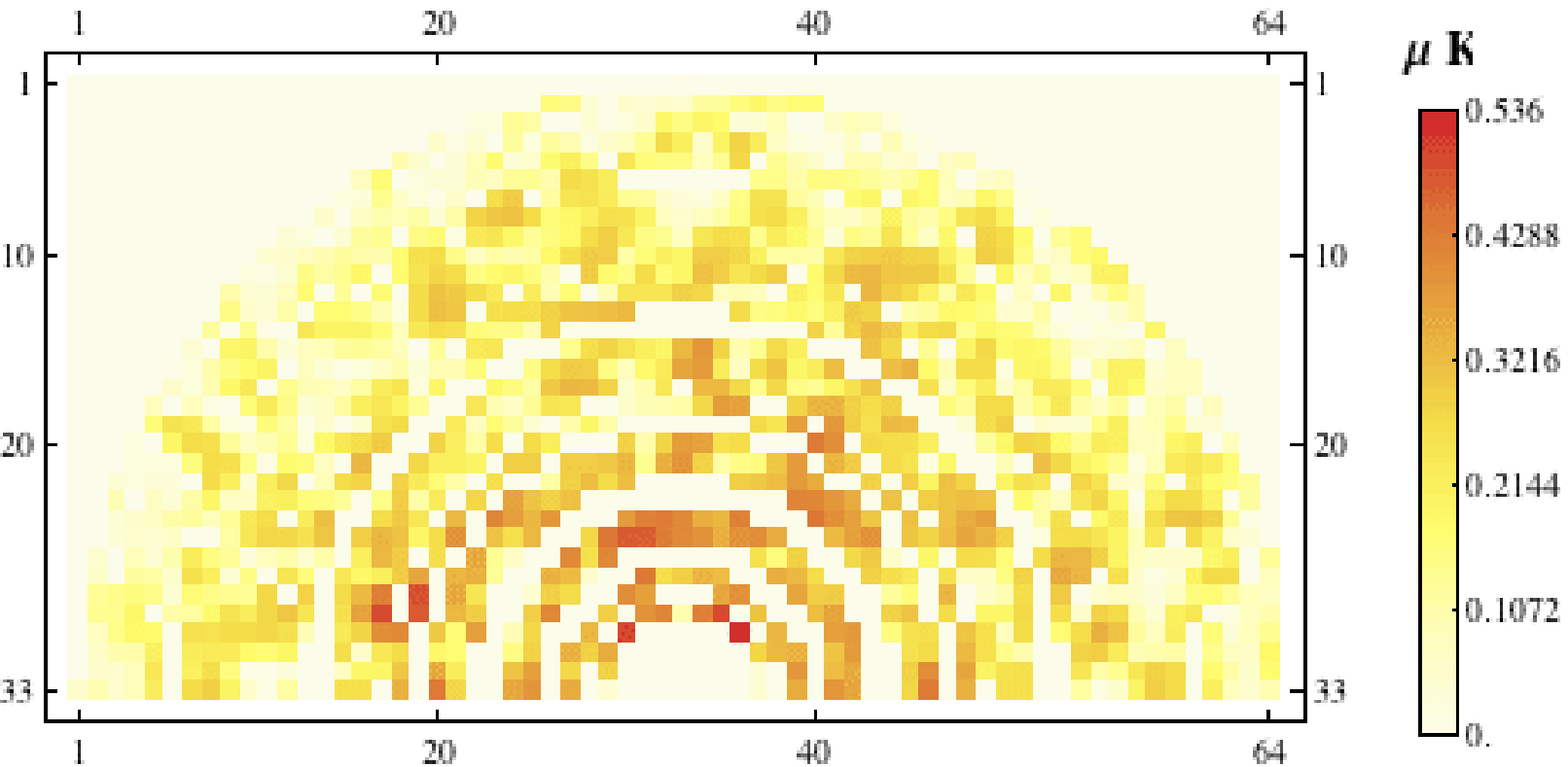} 
   \label{fig:uvI} }

 \subfigure[$V_Q({\bf u})$]{
   \includegraphics[scale =0.35] {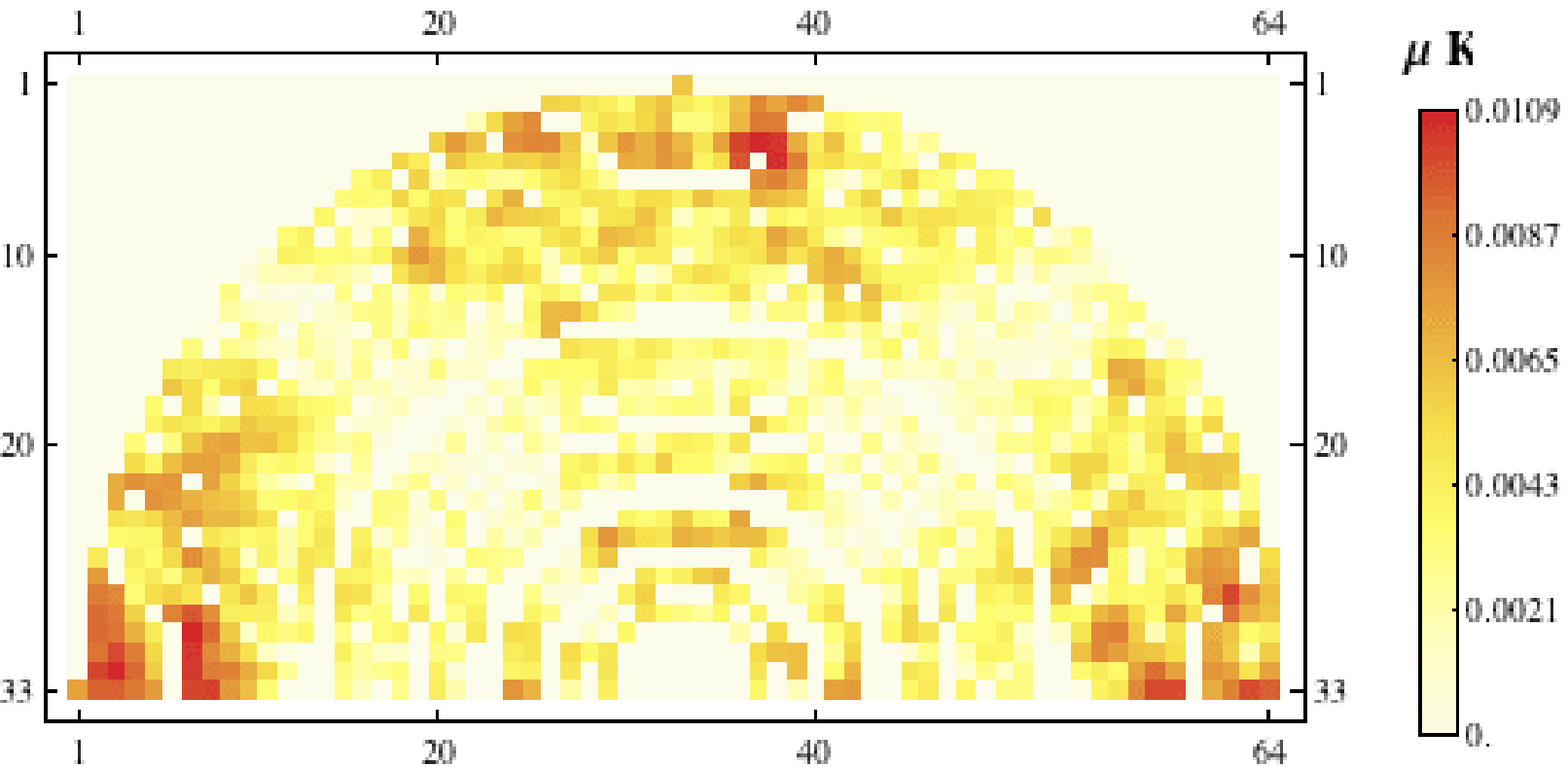}
   \label{fig:uvQ} }  

 \subfigure[$V_U({\bf u})$]{
   \includegraphics[scale =0.35] {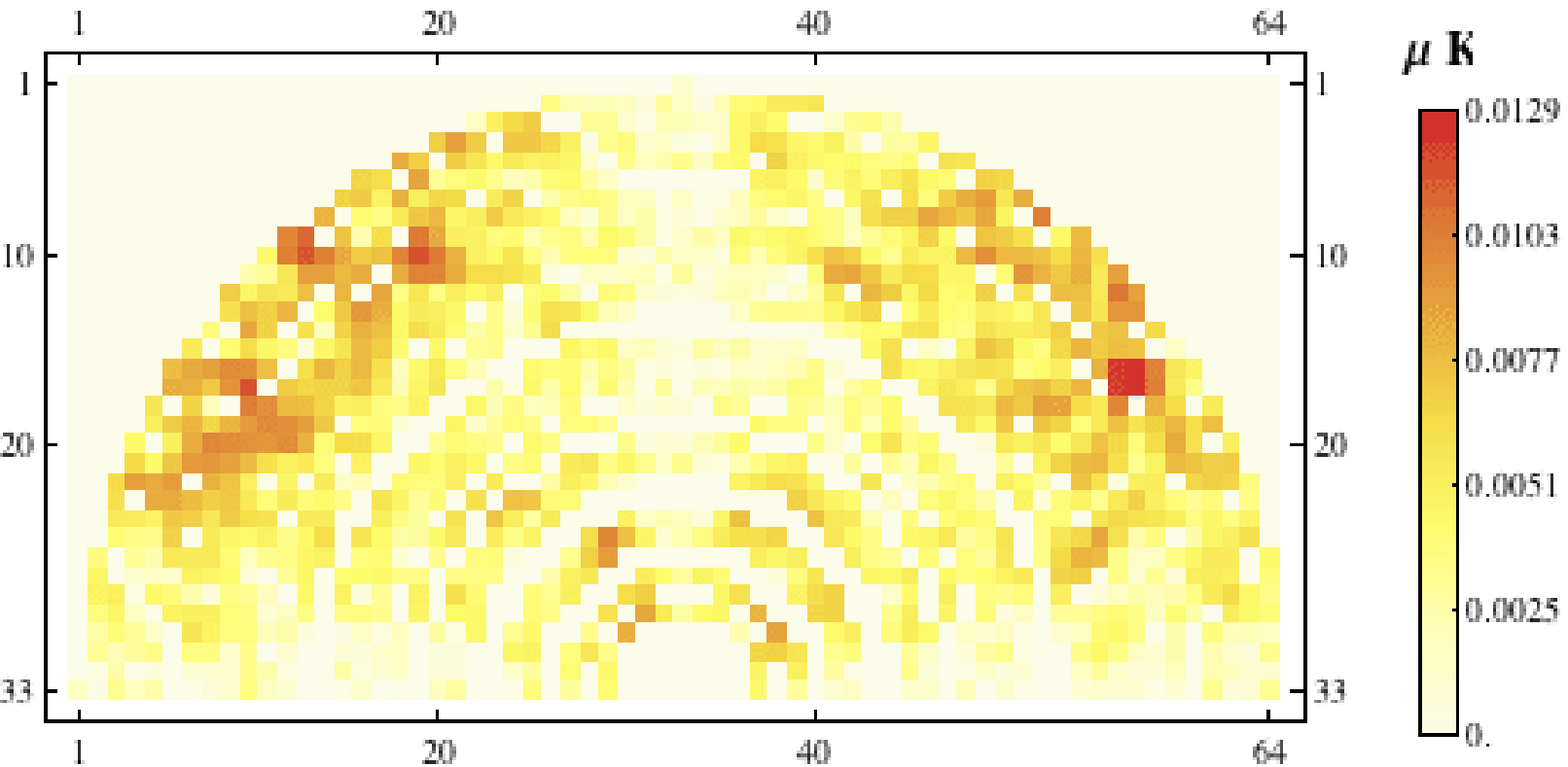}
   \label{fig:uvU} }
}

\label{fig:simIQU}
\caption{Simulated interferometric observations. The images shown in panels (a), (b) and (c) are a $30\times30$-degree realization of the two-dimensional CMB Stokes fields $I$, $Q$ and $U$ based on the standard CMB power spectra, with a $64\times64$ pixel grid.  All Fourier modes higher than the Nyquist frequency are filtered out to avoid aliasing. The images shown in the remaining panels are simulated Stokes visibilities (shown as magnitudes) by a QUBIC-like observation, assuming a Gaussian primary beam $A({\bf x})$ with beam width $\sigma=5^\circ$, a simulated 12-h $uv$-coverage of single field with $400$ close-packed antennas and Gaussian random noise of $0.015\mu$K per visibility. The map units are equivalent thermodynamic temperature in $\mu$K.}
\end{figure*}

We generate realizations of Stokes parameter maps having a physical size of 30 degrees on a side and resolution of $64\times64$ pixels. This large patch size ensures that the intensity beam pattern $|G|^2$ at the edges decreases to $\sim1$-percent level of its peak value. Although this size of patch seems to severely violate the flat-sky approximation, the primary beam pattern itself is small enough (the field-of-view $\Omega$ is about 0.047 sr) so that the flat-sky approximation is still valid. For simplicity, we assume that all the antennas continuously observe the same sky patch at a celestial pole and the interferometer is located at the north or south pole , the $uv$-tracks should be perfectly circular for a 12-h observation. Fig.~\ref{fig:simIQU} shows the mock systematics-free visibility data from these observations.

\section{Maximum Likelihood Analysis}
\label{max}
The maximum likelihood estimator of the power spectrum has many desirable properties~\citep{Bond:1998,Kendall1987}. The idea is to choose a model for the data and construct a likelihood estimator to evaluate how well the model matches the data. For a given model, comparing to the actual data set will give a likelihood of the model parameters. In practice, it is easier to maximize the logarithm of the likelihood function than the likelihood function itself.

Since CMB Stokes visibilities are complex Gaussian random variables with zero mean and dispersion $C_{V}+C_N$, the logarithm of the likelihood function is given by
\begin{equation} 
\ln\mathcal{L}(C_\ell) = n\log\pi-\log|C_V+C_N| - {\bf V}^\dagger(C_V+C_N)^{-1}{\bf V}  \, ,
\label{eq:loglike} 
\end{equation}
where ${\bf V}$ is the visibility data vector, $C_V$ is the signal covariance matrix predicted by $\langle{\bf V}^\dagger {\bf V}\rangle$, which can be constructed through Eq.~\ref{eq:eqcz}, and $C_N$ is the noise covariance matrix, which can be computed by Eq.~\ref{eq:CN}.

In practice, we parameterize the CMB power spectrum $C_\ell$ as flat band-powers over some multipole range to evaluate the likelihood function~\citep{Bunn:1996,Bond:1998,Gorski:1996,White1999}. We divide the power spectrum $\ell(\ell+1)C_\ell$ into $N_b$  piecewise-constant bins. Each bin corresponds to separate annuli in the $uv$-plane, characterizing the averaged $C_\ell$ over its bin-width. In our case, we evaluate the likelihood function by varying the CMB band-powers $\{\overline{\mathcal{C}_b^{TT}},\overline{\mathcal{C}_b^{EE}},\overline{\mathcal{C}_b^{BB}},\overline{\mathcal{C}_b^{TE}},\overline{\mathcal{C}_b^{TB}},\overline{\mathcal{C}_b^{EB}}\}$ with $b=1,\dotsc,N_b$. Here $\overline{\mathcal{C}_b}\equiv 2\pi|{\bf u}_b|^2S(|{\bf u}_b|)$. 

The bin-width can be chosen arbitrarily, but an appropriate choice of width is fine enough resolution to accurately detect the structure of the power spectrum and also wide enough to reduce the correlation between the band-power estimates so that the statistical errors on different band-power bins are approximately uncorrelated. 
The natural choice of bin-width can be approximated by the characteristic width of the Fourier transformed intensity beam pattern $A({\bf x})$, which defines the typical correlation length in the $uv$-plane. The minimum bin-width for the QUBIC-like experiment is about $\Delta u \approx\sqrt{\Omega^{-1}}=4.1$ wavelengths, corresponding to $\Delta \ell \approx26$. As a consequence, the total number of band-power bins is $6\times N_b\approx 84$.  However, the computational time required to evaluate the likelihood function in such a large number of bins is unfeasible.  Instead, in this paper, we estimate the power spectrum by using the bin-width of $\Delta\ell\simeq60$, roughly having 6 band-power bins for each power spectrum at the range of $28<\ell<384$.      

Using the above parametrization and following the previous papers~\citep{Hobson2002,Park2003,White1999}, the covariance matrices defined in Eq.~\ref{eq:eqcz} can be written as  
\begin{equation}
C^{ij}_{ZZ'} = \sum_{b=1}^{N_b} \sum_{\alpha,\beta} \overline{\mathcal{C}_b^{\alpha\beta}}  \int _{|{\bf u}_{b1}|}^{|{\bf u}_{b2}|}\, \frac{1}{2\pi}\frac{dw}{w} \times  W^{i,j}_{ZZ'\alpha\beta}(w) \, ,
\label{eq:eqczb}
\end{equation}
where we introduced the so-called {\em window functions} $W^{ij}_{ZZ'\alpha\beta}$ given by
 
\begin{equation} 
W_{ZZ'\alpha\beta}^{ij}(|{\bf w}|) = \int_0^{2\pi}d\phi_{\bf w}\, \omega_{Z\alpha} \omega_{Z'\beta}   \tilde{A}({\bf u}_i-{\bf w})\tilde{A}^*({\bf u}_j-{\bf w}) \, ,
\label{eq:wf} 
\end{equation}
where $Z,Z' = \{I,Q,U\}$ and $\alpha,\beta =\{T,E,B\}$ with $\omega_{IT}=1$, $\omega_{UE} = \sin 2\phi_{\bf w}$, $\omega_{UB} = \cos 2\phi_{\bf w}$, $\omega_{QE} = \cos 2\phi_{\bf w}$, $\omega_{QB} = -\sin 2\phi_{\bf w}$ and otherwise zero. 

Here we should note that the window functions $W_{ZZ'\alpha\beta}^{ij}(|{\bf w}|)$ are independent of the band-power spectra and therefore we only need to pre-calculate the integrals of the window functions over $w$ in Eq.~\ref{eq:eqczb} once for evaluating the covariance matrices.  Furthermore, if the primary beam pattern is Gaussian, the window functions can be integrated out analytically. Following~\cite{Hobson2002}, in Table~\ref{tab:windows} we provide the formulas of integrals for computing the window functions in Eq.~\ref{eq:wf}. This is all the formalism required for constructing the covariance matrices.

Empirically, direct evaluation of the full log-likelihood function over high-dimensional parameter spaces is unachievable. However, thanks to sophisticated and efficient numerical algorithms, it becomes possible to find the parameter values that maximize the log-likelihood function in relatively few steps, of order of $N_b^2$. As mentioned by~\cite{Hobson2002} and references therein, the most efficient numerical algorithm for maximizing the likelihood function is the combination of the {\em sparse matrix conjugate-gradient} algorithm and {\em Powell's directional-set} method. Due to the high sparsity of the covariance matrix for interferometer data, sparse matrix algorithms can dramatically reduce the computational time, by a factor of $f_s^{1.5}$ compared to the standard dense matrix algorithm, where $f_s$ is defined by the sparsity fraction of the covariance matrix.  Using Powell's direction-set method instead of the standard Newton-Raphson method, which requires intensive computations on the gradient or curvature of the log-likelihood function, the maximising process requires only $\sim3N^2$ function calls by the line-minimisation method. For a QUBIC-like observation of a single field, with 6 spectral bins in each CMB power spectrum, the maximum-likelihood evaluations for a total about 4000 visibilities in the $I,Q,U$ maps require about 20 hours of CPU time. 

Assuming that the likelihood function near its peak $\hat{{\bf a}}$ can be well approximated by a Gaussian, the parameter confidence intervals can be estimated by taking the inverse of the Hessian matrix ${\bf H}(\hat{{\bf a}})$, which is the matrix of second derivatives of the log-likelihood function with respect to the parameters, i.e. $\partial^2{\bf H}/{ \partial a_i \partial a_j}$.  The inverse of the Hessian matrix can be regarded as the asymptotic covariance matrix of the parameter estimates. The square roots of the diagonal elements of the asymptotic covariance matrix are assumed to be asymptotic standard errors of the parameter estimates, namely $\left< \delta a^2_i\right> = ({\bf H}^{-1})_{ii}$. Practically, we perform second differences numerically along each parameter direction to directly obtain the Hessian matrix and then calculate the {\em statistical error} estimates on the band-power spectra. We find this procedure requires only about 30 mins of CPU time for $\sim4000$ visibilities.     

\begin{figure*}[ht]
\centering
\mbox{
\subfigure[$TT$ power spectrum]{
   \includegraphics[scale =0.27] {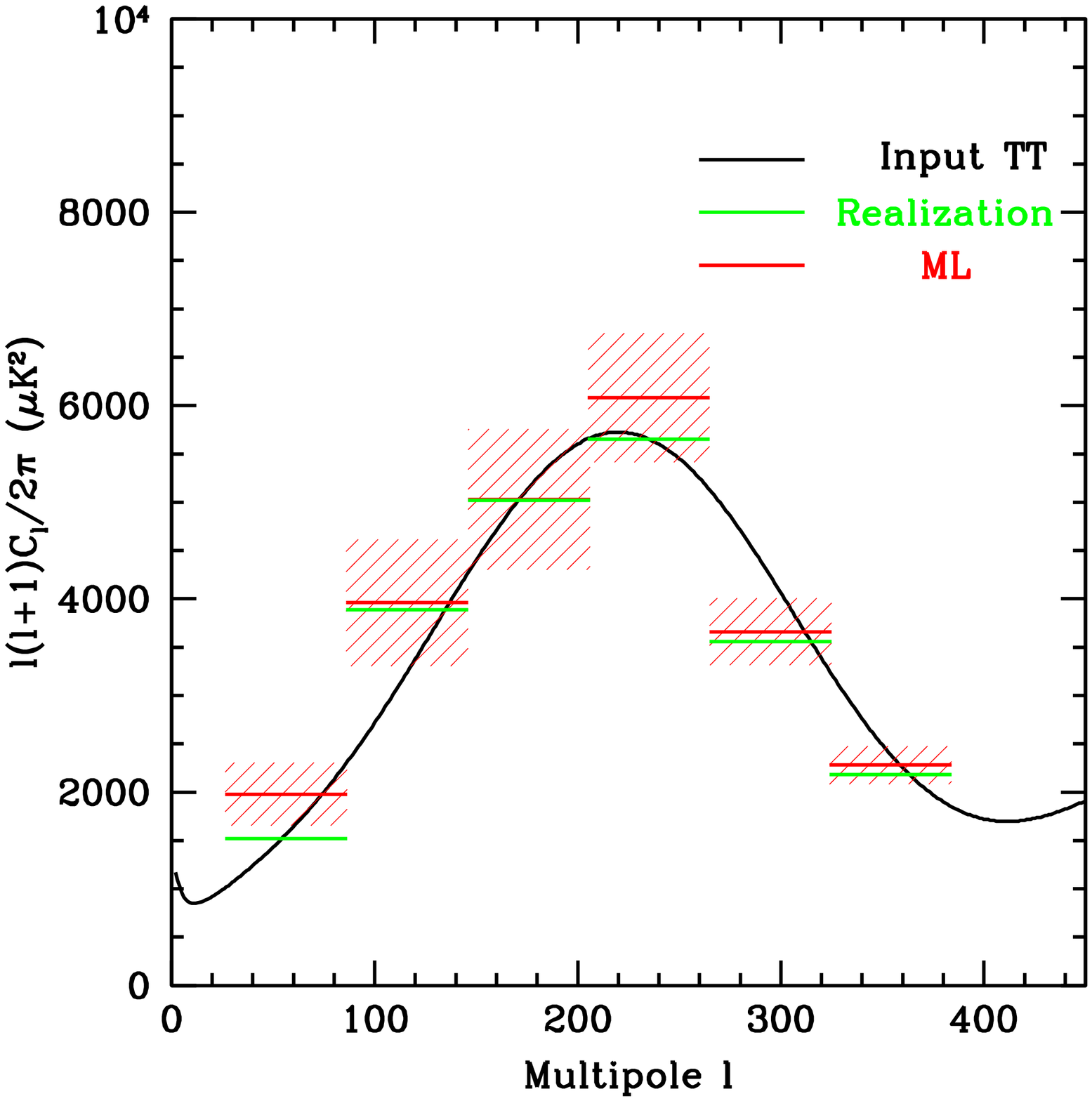} 
    }

 \subfigure[$EE$ power spectrum ]{
   \includegraphics[scale =0.27] {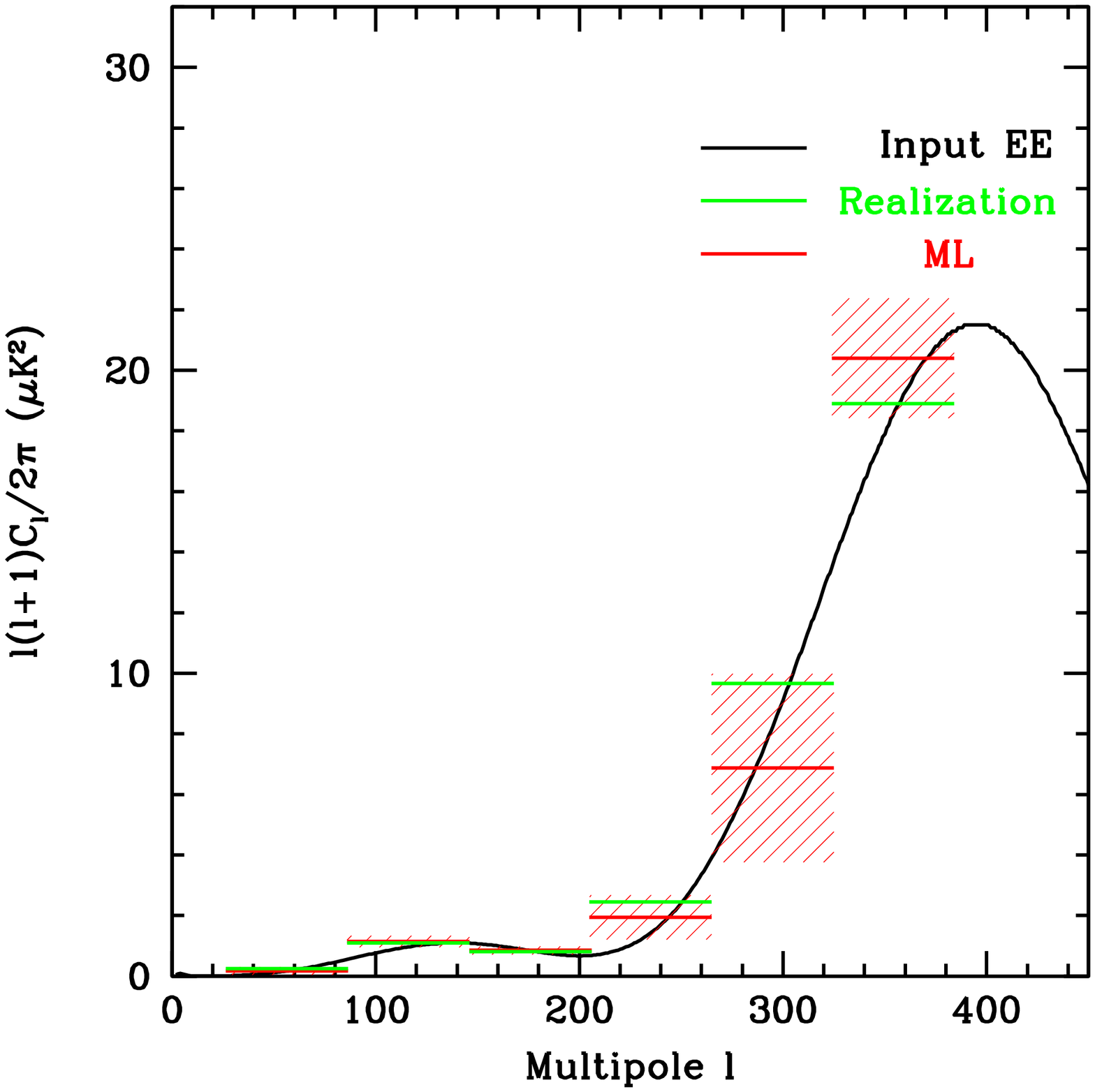}
    }  

 \subfigure[$BB$ power spectrum]{
   \includegraphics[scale =0.27] {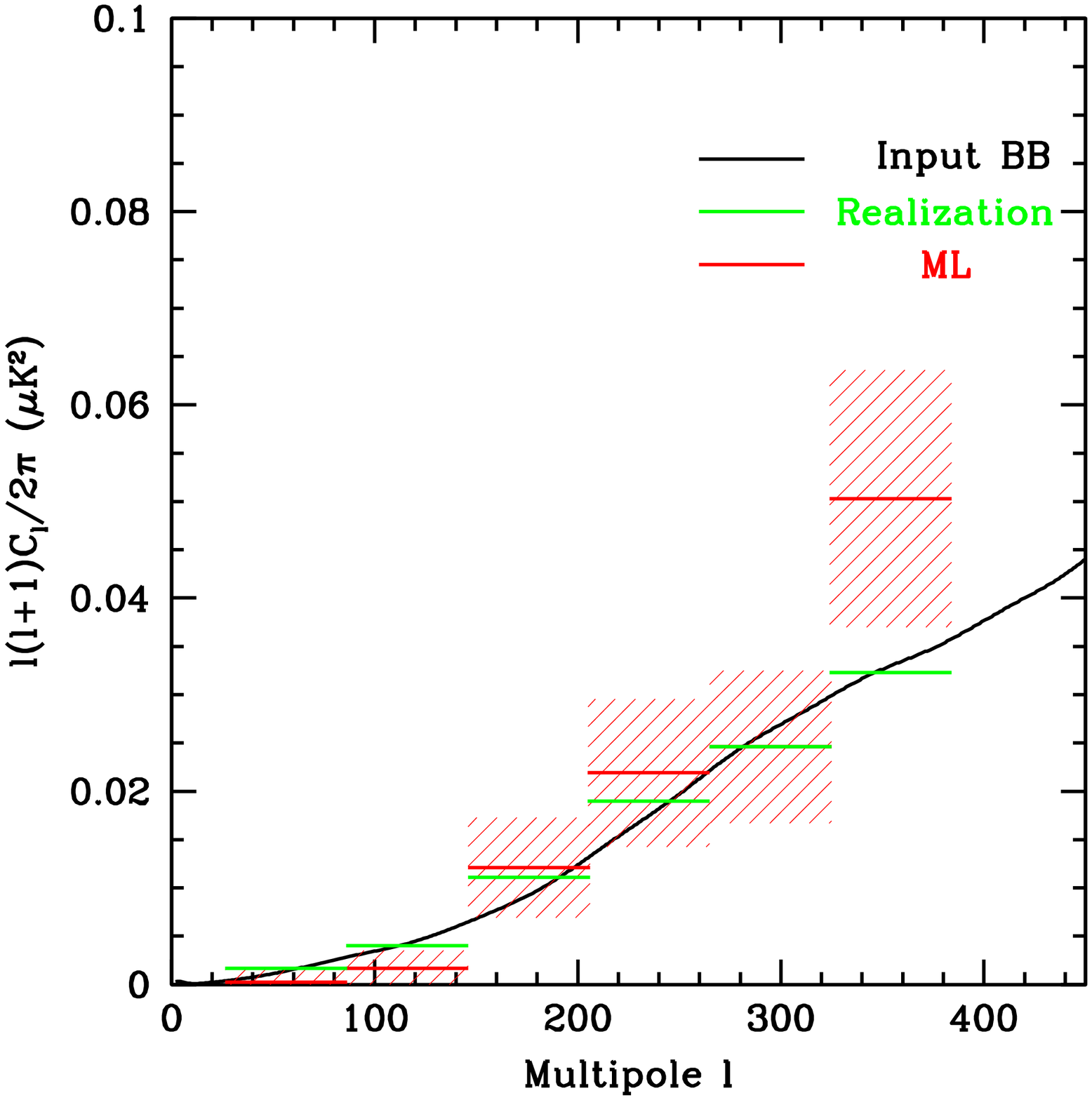}
   }
}

\mbox{
\subfigure[$TE$ power spectrum]{
   \includegraphics[scale =0.27] {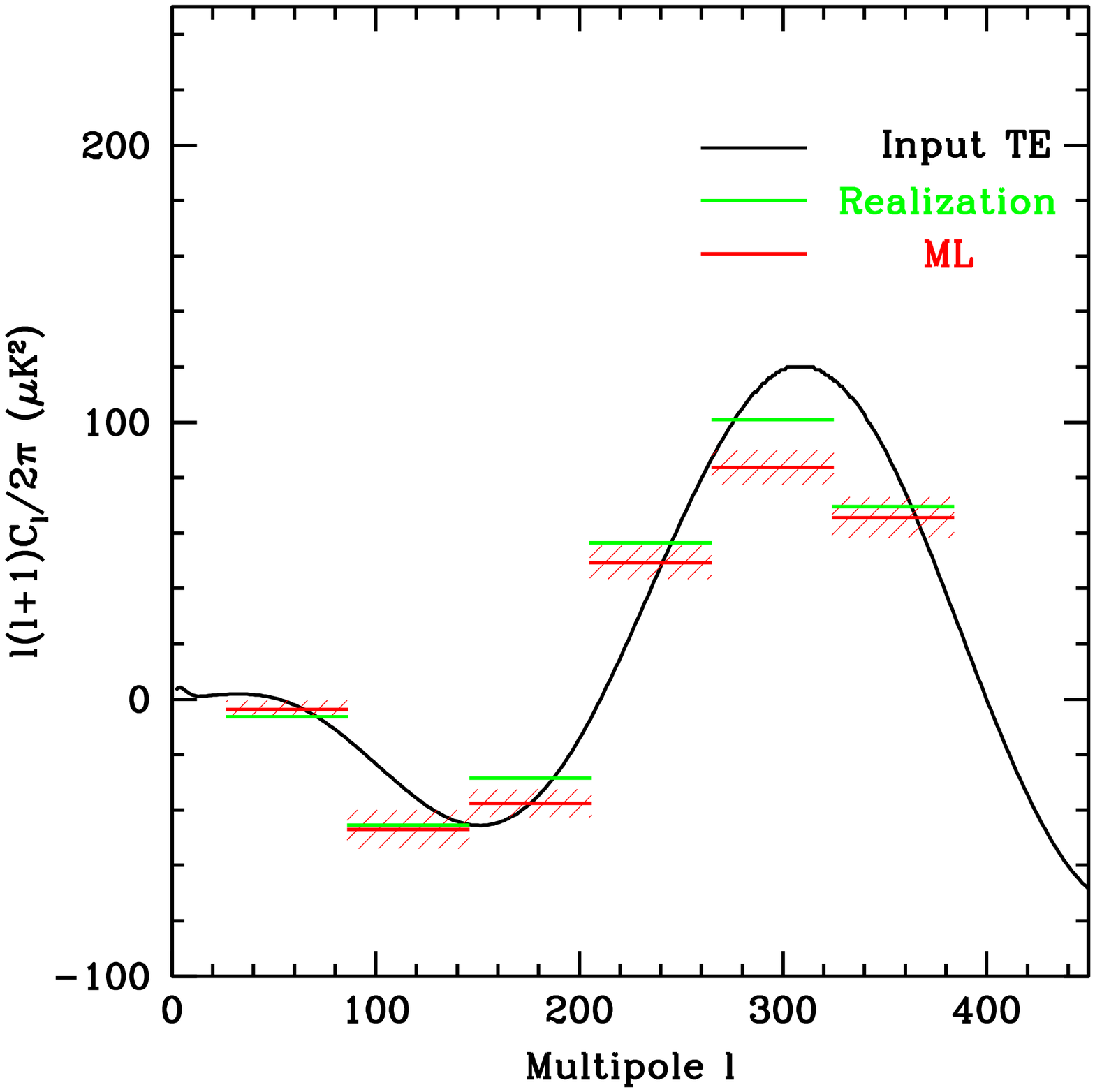} 
   }

 \subfigure[$TB$ power spectrum]{
   \includegraphics[scale =0.27] {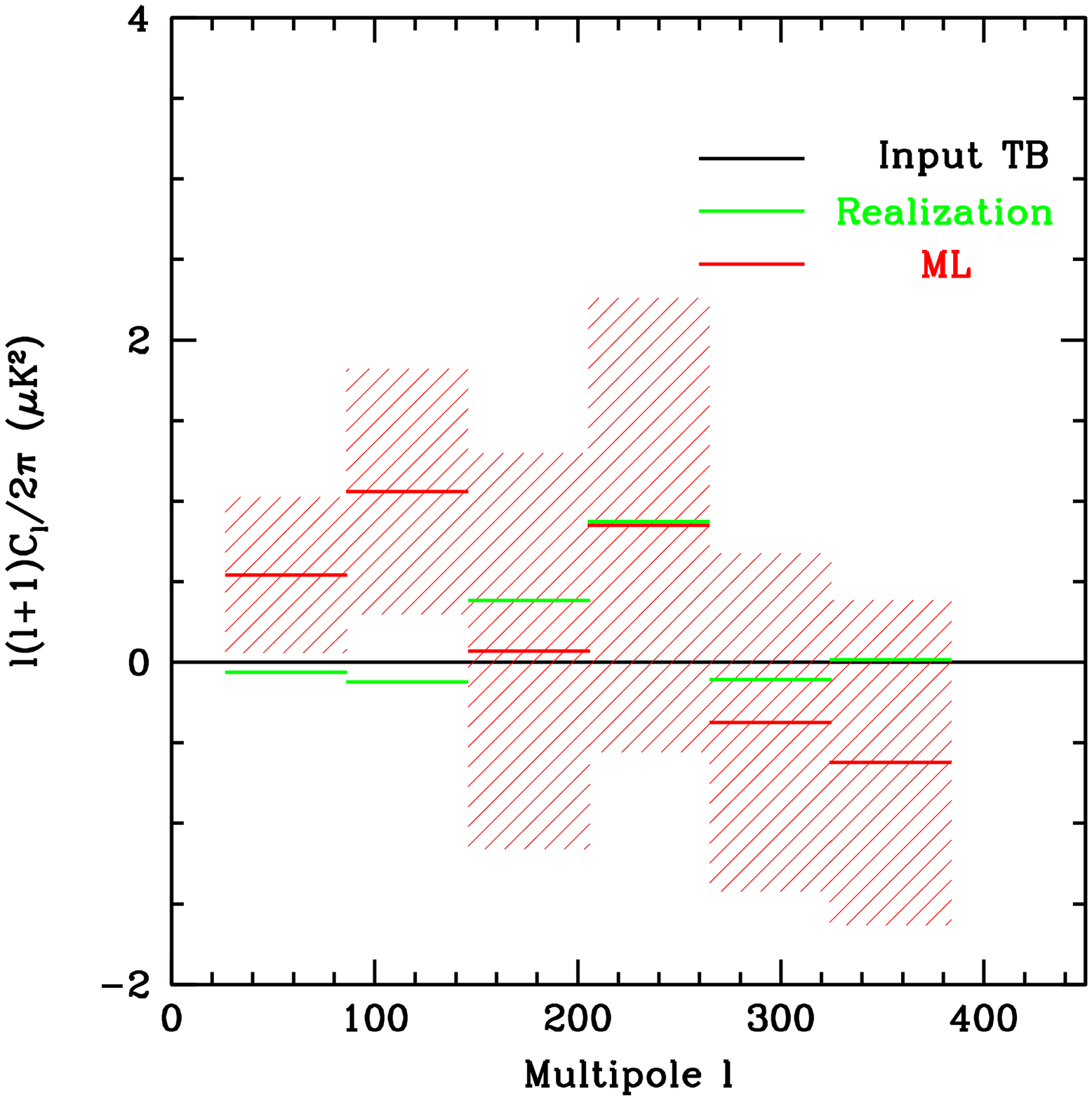}
   }  

 \subfigure[$EB$ power spectrum]{
   \includegraphics[scale =0.27] {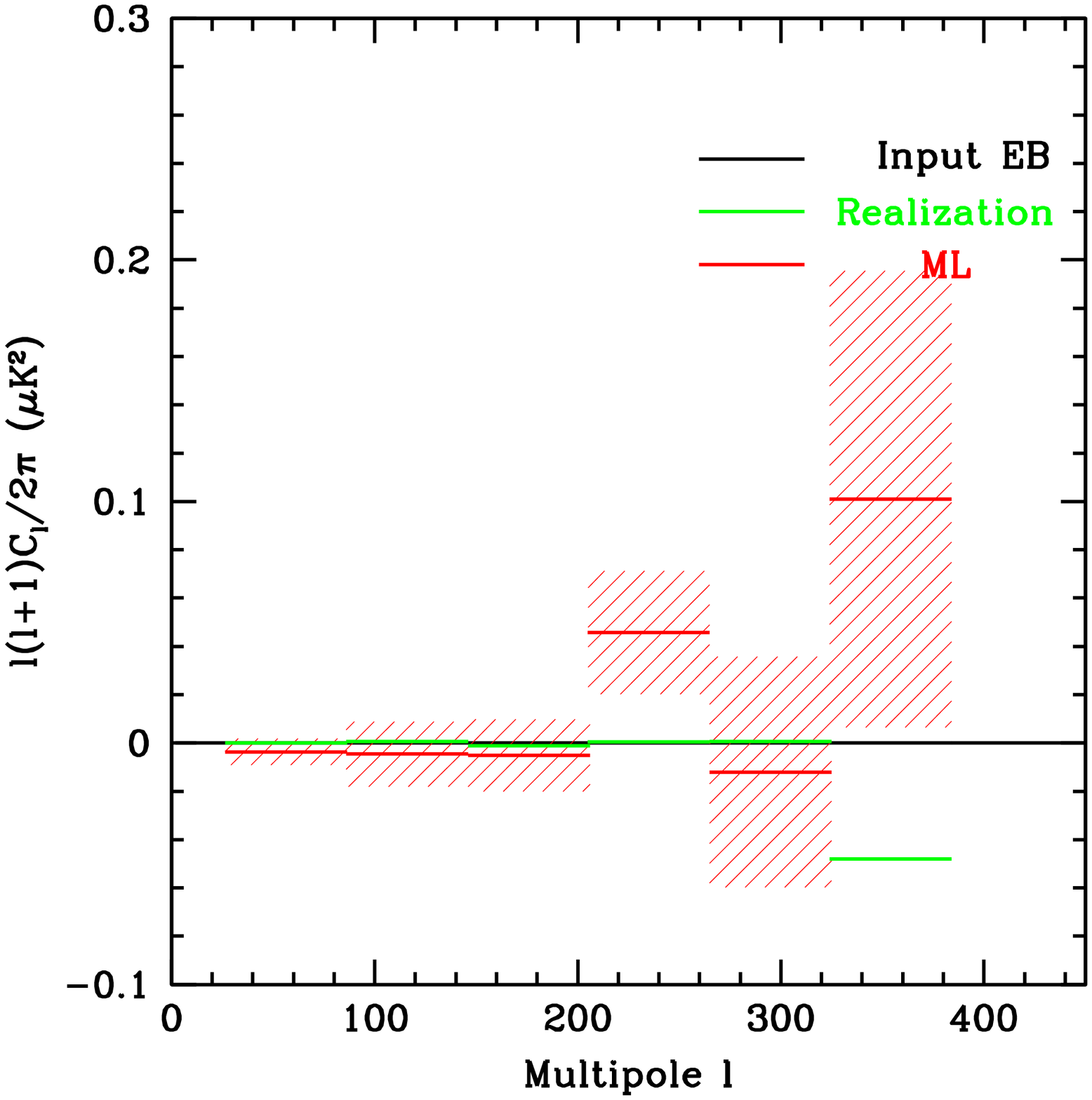}
    }
}

\caption{The CMB power spectra $TT,EE,BB,TE,TB,EB$ (red) recovered by maximizing the likelihood function from a mock QUBIC-like observation in the absence of systematic errors (see details in text). The flat band powers are estimated in 6 bins with bin-widths $\Delta u= 9.5$ and with 1-$\sigma$ (red hatched) statistical uncertainties. The input CMB power spectra are shown in black and the specific realizations of such input power spectra are shown in green.}
\label{fig:tcl}
\end{figure*}

Fig.~\ref{fig:tcl} shows the resulting maximum-likelihood CMB power spectra based on the simulated observations in the absence of systematic errors. The recovered power spectra are basically consistent with the true underlying CMB power spectra within 2-$\sigma$.

\section{Analysis of pointing errors}
\label{res}
Using the simulated Stokes visibilities and applying the ML analysis described in the previous section, we obtained estimates of the systematic pointing errors in the CMB power spectra. Other systematics, such as beam shape errors, gain errors and cross polarization, will be presented in a detailed analysis on various instrumental systematics in a forthcoming paper.  

The quadrature difference between the recovered power spectrum  with and without pointing errors is used to estimate the effect of systematic pointing errors. Also the pointing errors can potentially change the statistical error (which depends on the curvature of the likelihood function) for a given experiment. The bias in the $i$-th band-power spectrum $\overline{\mathcal{C}_i}$ and the change in  the corresponding statistical error $\overline{\sigma_i}$ are given by  
\begin{eqnarray}
&& \Delta\overline{\mathcal{C}_i} = \langle( \overline{\mathcal{C}_i^{error}} -  \overline{\mathcal{C}_i})^2\rangle^{1/2} \nonumber \\
&& \Delta \overline{\sigma_i} = \langle(\overline{\sigma^{error}_i} - \overline{\sigma_i})^2\rangle^{1/2} \, , 
\label{eq:dcl}
\end{eqnarray}
where $\overline{\mathcal{C}_i^{error}}$ refers to values obtained in the presence of pointing errors and $\overline{\mathcal{C}_i}$ is the recovered power spectrum for that same patch of the sky in the absence of systematic errors (refers to ``ML'' in Fig.~\ref{fig:tcl}).  To quantify how significant this systematic error is when compared with the systematics-free 1-$\sigma$ statistical error, following~\cite{Dea:2007} and~\cite{Miller:2008}, we introduce the tolerance parameters defined by 
\begin{eqnarray}
&& \alpha_i= \frac{\Delta \overline{\mathcal{C}_i}}{\overline{\sigma_i}}  \nonumber\\
&& \beta_i= \frac{\Delta \overline{\sigma_i} }{\overline{\sigma_i}}\, ,
\label{eq:torl}
\end{eqnarray}
where $\Delta \overline{\mathcal{C}_i} $ and $\Delta \overline{\sigma_i} $ are quadrature differences in Eq.~\ref{eq:dcl}. We set up a tolerance limit, say $10\%$, which requires that neither $\alpha$ nor $\beta$ exceed $0.1$. In our simulations, we consider two types of pointing errors, which we call {\em uncorrelated} and {\em fully-correlated}. In the uncorrelated case, the pointing errors are assumed to be Gaussian-distributed for each antenna (and tend to average out), and in the fully-correlated case, the pointing offsets $\Delta {\bf x}$ relative to the desired observing direction are identical for all the antennas and remain fixed on the sky as the sky rotates. Without pointing errors, the desired observing direction is fixed to a celestial pole where all the antennas continuously observe the same sky patch as the sky rotates. Once the pointing errors are present in real observations, the ``actual'' observing direction for each antenna would change with time as the sky rotates since the direction being observed in a realistic scenario is fixed with respect to the Earth rather than to the sky. To exactly mimic the visibility signal measured by each antenna pair would require observing a patch that shifts with time as the sky rotates.  Further, the observed patches would be slightly different for each pair of antennas. For simplicity, we assume the pointing offsets remain fixed on the sky instead of with respect to the Earth as a good approximation if the pointing offsets are randomly Gaussian-distributed in each antenna and are small enough compared to the beam width. This approximation would be less valid in the fully-correlated case which corresponds to the entire array of antennas simply looking at the same ``wrong'' patch of sky, although it would be valid at any instant. We leave a more detailed investigation on the pointing effects taking into account the sky rotation to a future paper. Our preliminary results from this more complete calculation show that the resulting biases in the $BB$ measurement are consistent within $\sim30\%$ in these two scenarios for the rms pointing errors of $0.7^\circ$.

\begin{figure*}[ht]
\centering

\mbox{
\subfigure[$BB$ power spectrum]{
   \includegraphics[scale =0.27] {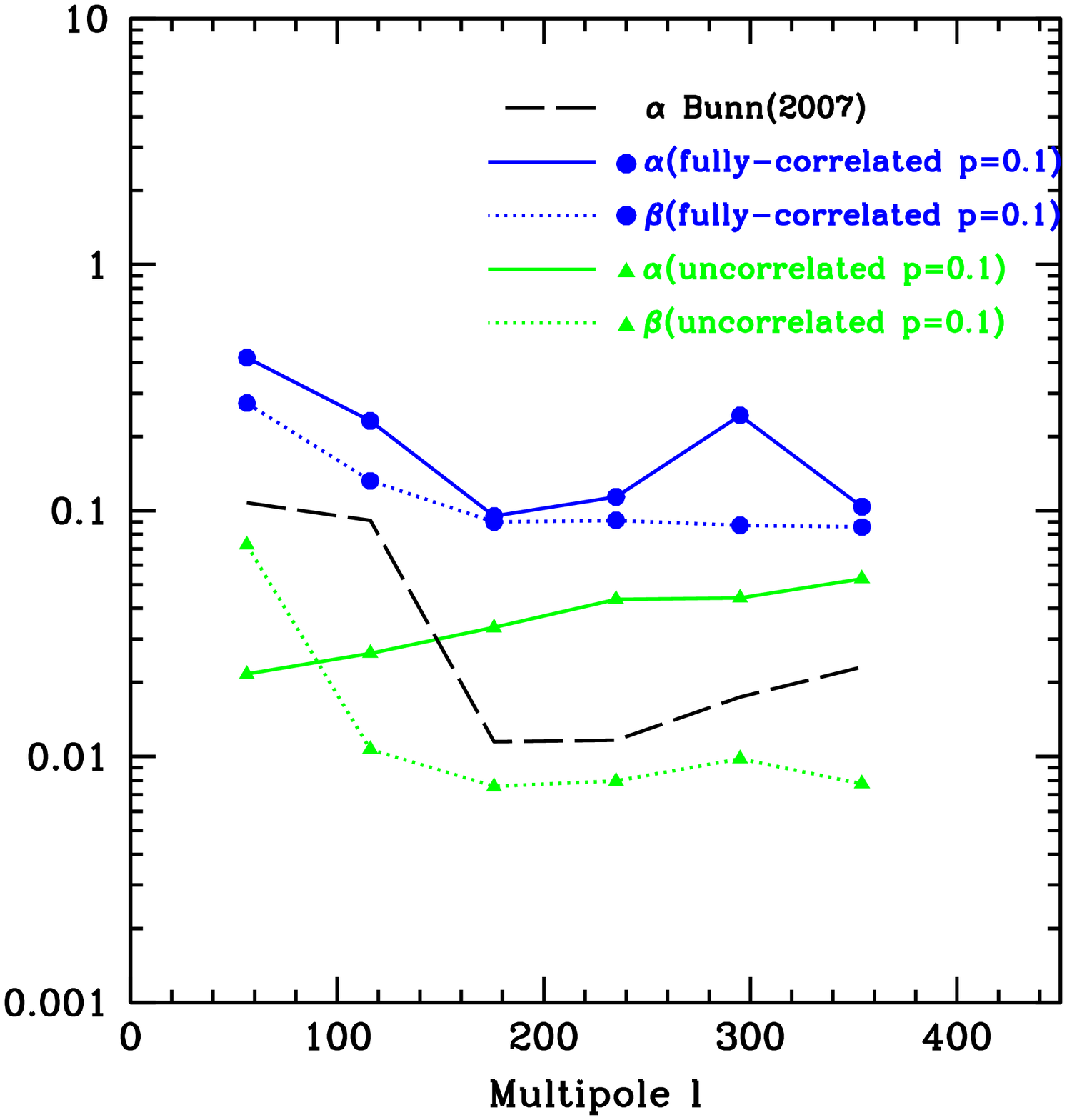}
   \label{fig:picBp} }

 \subfigure[$TB$ power spectrum ]{
   \includegraphics[scale =0.27] {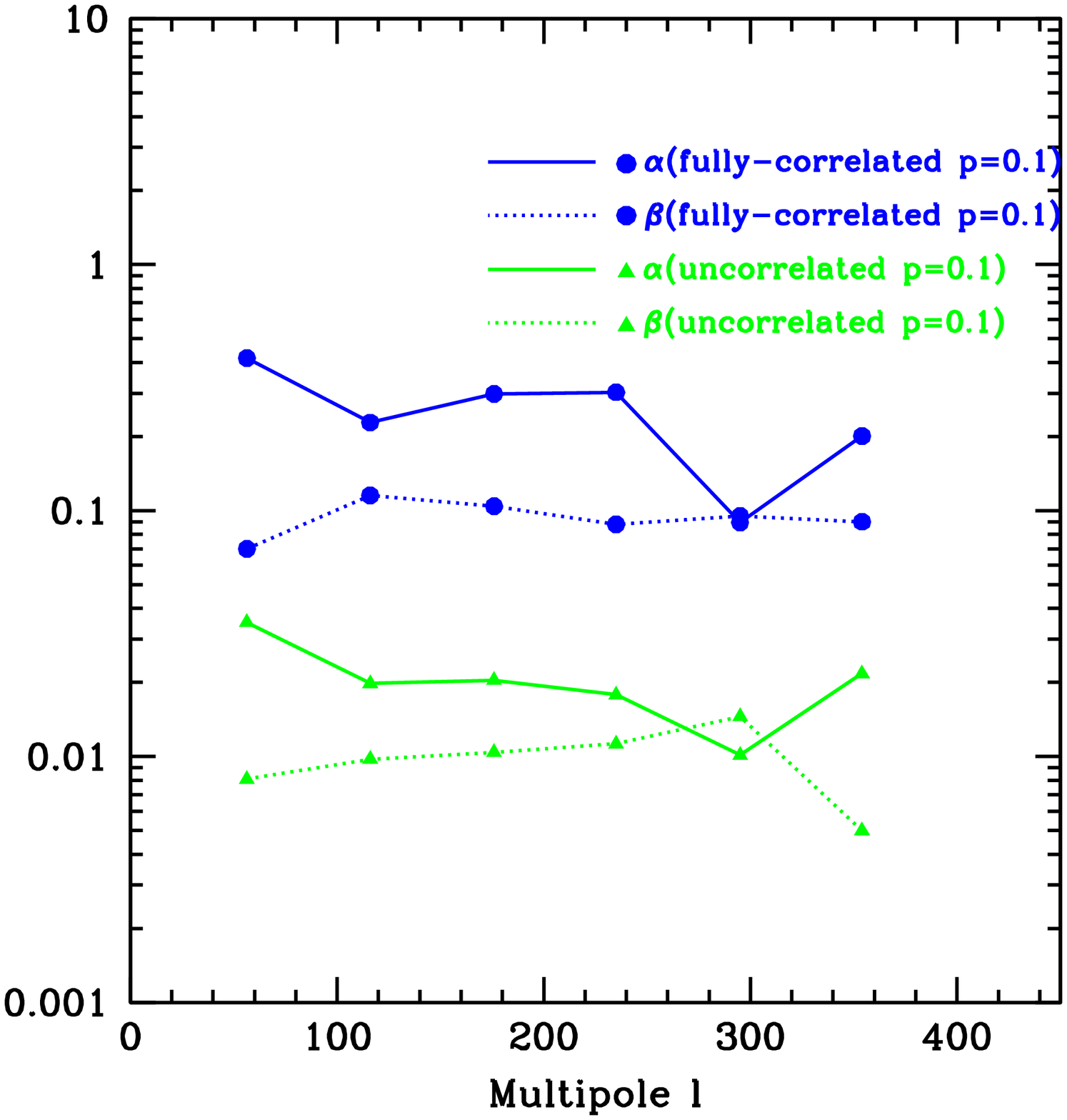}
   \label{fig:picTBp} }

 \subfigure[$EB$ power spectrum]{
   \includegraphics[scale =0.27] {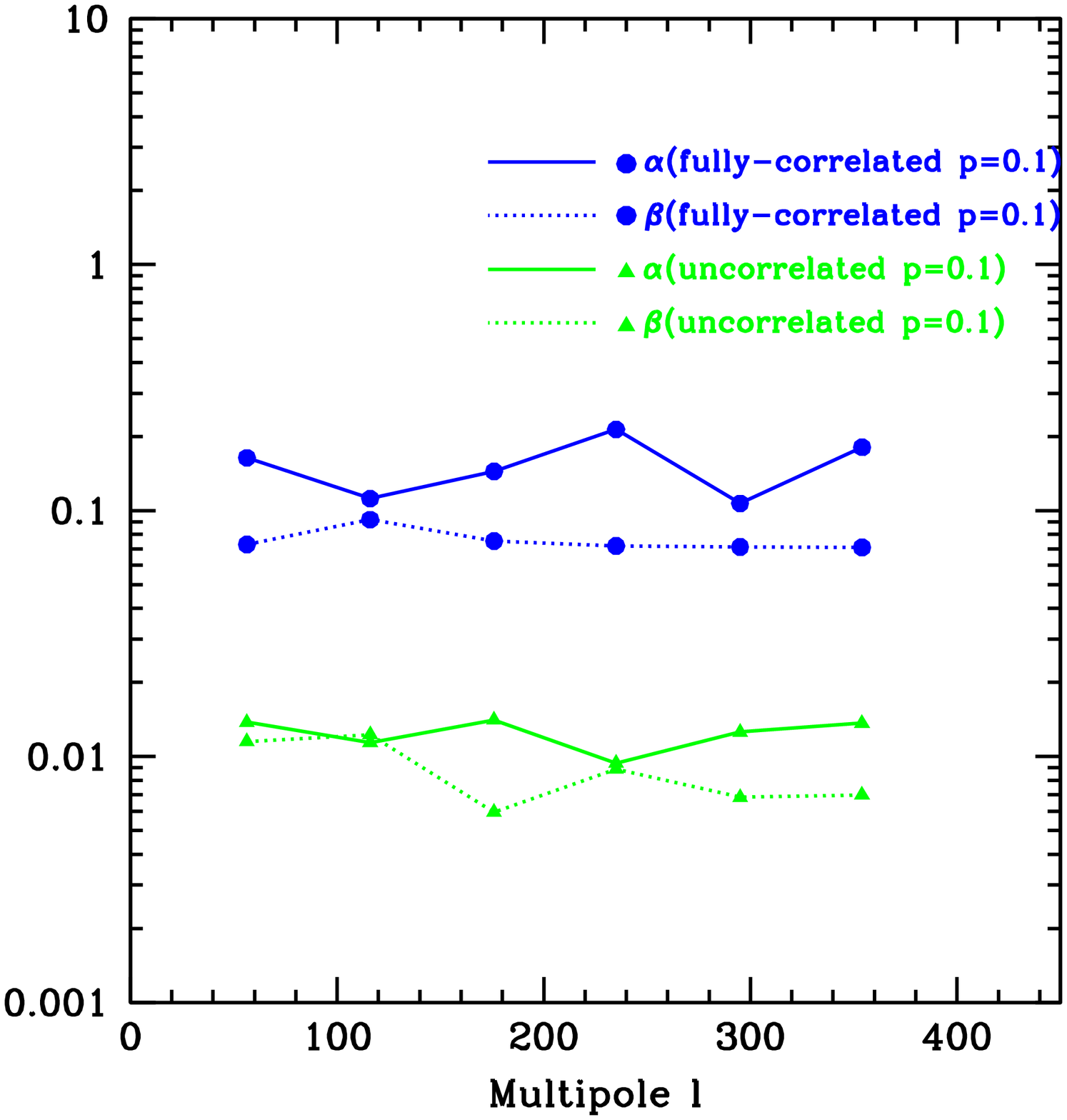}
   \label{fig:picEBp} }
}

\caption{Assessment of systematic pointing error for $BB$, $TB$ and $EB$ power spectra. In the units of systematics-free 1-$\sigma$ statistical uncertainties, the solid lines are for the induced bias in the power spectra and dotted for the changes in statistical uncertainties. Green triangles denote the results in the case of the independent Gaussian-distributed pointing errors with the dispersion $\delta=0.1\sigma$ for each antenna, which are estimated by averaging over 50 simulations. The corresponding analytical result by~\citet{bunnsyst} is shown in dashed-black line. Blue points correspond to all the antennas having the identical pointing offsets with the amplitude $|\Delta{\bf x}|= 0.1\sigma$ and the results are obtained by  averaging over 20 random directions relative to the target direction.}
\label{fig:pcl}
\end{figure*}

As introduced by~\cite{bunnsyst}, we use an error parameter $p$ in units of the beam width $\sigma$ to characterize the pointing error level, i.e., $p=\delta/\sigma$ and $p= |\Delta{\bf x}|/\sigma$ respectively for the uncorrelated and fully-correlated cases, where $\delta$ is the root-mean-square (rms) value of Gaussian-distributed pointing errors and $|\Delta{\bf x}|$ is the amplitude of identical pointing offsets.  For clear comparison, Fig.~\ref{fig:pcl} shows the values of $\alpha$ and $\beta$ in the $BB$, $TB$ and $EB$ power spectra with the  identical $p=0.1$ for both the cases. As illustrated in Fig.~\ref{fig:pcl} for the uncorrelated case, the parameter $\alpha$ in each band-power bin for the $BB$ power spectrum is smaller than $0.06$ and smaller than $0.04$ and $0.01$ for the $TB$ and $EB$ power spectra, respectively. Also, the parameter $\alpha$ for the $BB$ power spectrum increases approximately monotonically with increasing $\ell$. Moreover, it is apparent that for the $BB$ power spectrum, the values of the parameter $\beta$ are always much smaller than those of $\alpha$ over the whole multipole range, except for the lowest band-power bin where $\beta$ roughly approaches our criterion of $10\%$ threshold. But for the $TB$ and $EB$ power spectra, the values of $\beta$ demonstrate that the changes in statistical errors in all band-power bins are basically comparable and they have less sensitivity than $\alpha$. Using $\alpha$ and $\beta$, and taking the tolerance limit of $10\%$, we put an upper limit on the allowed range of the pointing error parameter $p$, requiring $p\lesssim0.1$.

Here $\alpha$ and $\beta$ are estimated by the mean of 50 simulations, which in principle should be an accurate estimation of the true systematic contamination. To check the convergence of our results, we repeated another independent 50 realizations and found the variations of both $\alpha$ and $\beta$ to be less than $5\%$.  In order to clearly illustrate the systematic effects, for each simulation we randomly generate Gaussian pointing offsets for all the antennas, but the Stokes parameters, $I({\bf x})$, $Q({\bf x})$ and $U({\bf x })$, and the noise in each visibility are fixed during the entire simulation. In fact, our results are insensitive to different realizations of the sky maps and noise and only depend on the amplitude of the pointing error.

To verify our results we compare them with the analytical results by~\cite{bunnsyst}. In our simulations, we assume two visibilities $Q$ and $U$ are measured with the same antenna (as is the case of a circular experiment in~\cite{bunnsyst}), and therefore using the corresponding relation, the bias on the $B$-mode power spectrum is 
\begin{equation}
(\Delta C_\ell)^2 = \frac{p^2}{N}\left(8(\overline{s^2})^2(C_\ell^{EE})^2 + 6\overline{s^2}C_\ell^{EE}C_\ell^{BB} \right)\, , 
\label{eq:tedan}
\end{equation}
where $\bar{s^2}$ is the average of $\sin^2{2\phi}$ over the antenna patterns and the factor $N$ is the number of baselines contributing to each band-power bin since the systematic effects in random pointing offsets will average down as $\sqrt{N}$. The value of $\bar{s^2}$ approximately follows $\bar{s^2} =262.7/\ell^2$ in a QUBIC-like experiment and our simulation shows that the number of the baselines in the first two bins is about 2000 and about 5000 in the remaining bins. 
Due to pointing errors mostly arising from mixing $EE$ into $BB$, we only consider the leading contribution from $C_\ell^{EE}$ and the secondary one from $C_\ell^{EE}$ and $C_\ell^{BB}$. Using the above parameters, the corresponding results are shown in Fig.~\ref{fig:picBp}, from which we find that the simulation-based results are consistent within a factor of 4 with the analytical-based results. The simulated observations with likelihood-function analysis therefore provide reliable estimates on pointing errors.  Furthermore, numerical simulations illustrate that the analytical-based results by~\cite{bunnsyst} actually underestimate the biases at $\ell\gtrsim150$, since the analytical calculation can only give a first order approximation and thus provide lower bounds on pointing errors, which can yield poor estimates in cases of large pointing errors. Therefore the analytical-based results can be considered as an approximate estimation of systematic effects. The resulting bias that the simulations predict could be larger than those found by the analytical calculations.

In the fully-correlated case, the pointing directions of all antennas are offset by the same amount, which means they all look at a slightly different patch of the sky from the one used to compute $\overline{\mathcal{C}_i}$ in Eq.~\ref{eq:torl}. Recall $\overline{\mathcal{C}_i}$ is the recovered band-power spectrum in the absence of systematic errors. In this case, the systematic errors would be equivalent to the cosmic variance and $\alpha\simeq 1$ if instrumental noises were zero. It is worth noticing that the fully-correlated case would have no effect on the determination of cosmological parameters since observing a different patch of the statistically isotropic CMB sky would not affect the correct estimates of the underlying CMB power spectra. The simulations confirm, as expected, that the contamination levels in $\alpha$ and $\beta$ in each band-power bin are much larger than in the case of uncorrelated errors by a factor of $\sim10$. In the fully-correlated case pointing errors of $10\%$ beam width can affect the $BB$ and $TB$ power spectrum measurements at roughly the $50\%$ level in the lowest $\ell$-bin. Similar to the uncorrelated case, the parameter $\alpha$ is more sensitive to the pointing errors than $\beta$. The fully-correlated case however is a worst-case scenario and unrealistic. The actual pointing offsets should be very close to Gaussian distributions, which implies that the realistic contamination induced by pointing errors would resemble the uncorrelated pointing error forecasts. We conclude that for a QUBIC-like observation, the tolerance parameters $\alpha$ and $\beta$ remain below our tolerance limit of $10\%$ for pointing uncertainties as large as $p=0.1$ for all band powers. For the $EB$ and $TB$ power spectra, pointing errors are so small that we can entirely neglect such systematic effects in all band powers.

\begin{figure}[h]
  \centering
  \includegraphics[width=3.3in]{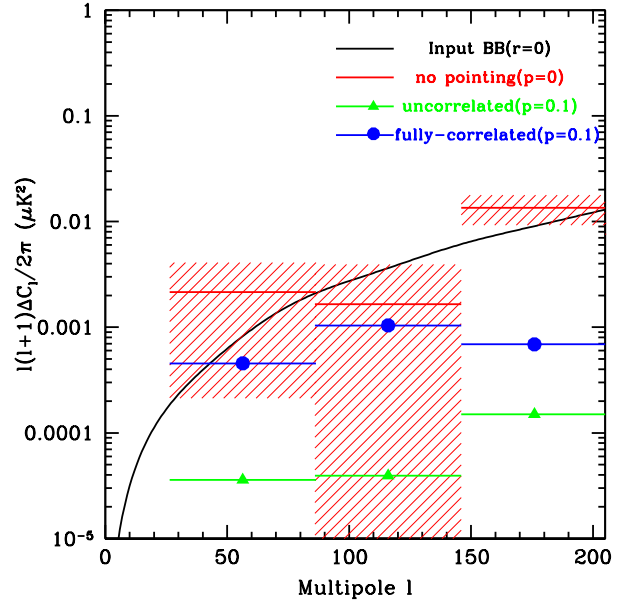}
  \caption[]
   {Overview of the effects of pointing errors on the B-mode power spectrum. The input $BB$ purely from gravitational lensing and not induced by primordial fluctuations is shown in black.  In the absence of systematic errors, the recovered $BB$ and the corresponding 1-$\sigma$ (red hatched) statistical uncertainty are also shown for comparison. The amplitudes of the false B-modes $\Delta \overline{\mathcal{C}}$ induced by pointing errors in the fully-correlated and uncorrelated cases are estimated by 50 realizations.}
\label{fig:deltacl}
\end{figure}

Furthermore, since a QUBIC-like interferometer is designed specifically to probe the primordial $B$-modes, we are mostly concerned with levels of bias in $r$. Due to the presence of sampling variance, instrumental noise and systematic errors, we can not perfectly recover the almost zero $BB$ in the lowest $\ell$-bin -- where the amplitude depends primarily on $r$ rather than on the lensing-induced signal if $r>0.01$ -- but can set an upper limit on $r$. We thus run simulations on the maps with the input $C^{BB}_{\ell}$ for $r=0$ to generate visibilities that include the effects of systematic pointing errors. Because the input $BB$ has $r=0$, any ``non-zero'' $BB$ signal in the lowest $\ell$-bin would lead to an upper limit on $r$, which is assumed to be a conservative estimate since the lensing-induced $B$-modes in the lowest $\ell$-bin can be partially removed using the signals in high-$\ell$ bins. The amplitude of the  false  $BB$ induced by pointing errors that couple $E$ to $B$ can be characterized by the quadrature difference $\Delta \overline{\mathcal{C}_i}$ between the recovered band-power with and without pointing errors, as in Eq.~\ref{eq:dcl}. By applying the ML approach to analyze the simulated signals from $r=0$ input $BB$,  we find that, in the absence of systematic errors, the amplitudes of the recovered $BB$ band-power in the lowest $\ell$-bin ($28<\ell<88$) and its 1-$\sigma$ statistical error are $\overline{\mathcal{C}}\simeq 2.1\times 10^{-3} \mu {\text K^2}$ and  $\overline{\sigma}\simeq1.9\times 10^{-3} \mu {\text K^2}$, respectively, as shown in Fig.~\ref{fig:deltacl}. Assuming that the theoretical primordial $BB$ band-power without lensing contributions in the lowest $\ell$-bin is $4.74\times10^{-4}\left(\frac{r}{0.01}\right) \mu{\text K^2}$, the amplitudes of the recovered $\overline{\mathcal{C}}$ and $\overline{\mathcal{\sigma}}$ are consequently comparable to the inflationary $BB$ power for $r=0.045$ and $\sigma_r=0.041$, yielding an upper limit of $r < 0.045+2\times0.041=0.127$ at $95\%$ confidence for systematic-free observations. In the presence of the pointing errors, the false $BB$ power in the uncorrelated case with $p=0.1$ has an amplitude of $3.6\times  10^{-5} \mu {\text K^2}$, which translates into an uncertainty in $r$, i.e., $\Delta r \simeq 7.6\times10^{-4}$. Moreover, in the fully-correlated case, the pointing-error-induced false $BB$ power is about 13 times larger than in the uncorrelated case, resulting in $\Delta r \simeq 0.01$. In addition,  the simulations show that the changes in the statistical errors  (as in Eq.~\ref{eq:dcl}) result in $\Delta \sigma_r\simeq 2.2\times10^{-3}$ and $5.8\times10^{-3}$ in the uncorrelated and fully-correlated cases, respectively. As a result, we can constrain $r<0.147$ and $r<0.132$ (i.e. $r< 0.127+\Delta r+2\times \Delta\sigma_r$)  in the fully-correlated and uncorrelated cases with $p=0.1$, respectively. Based on the above analysis, we can therefore conclude that, for a QUBIC-like observation of a single field, the pointing errors would slightly bias the 2-$\sigma$ upper limit of $r$ at $\sim 10\%$ level, which is much smaller than the statistical uncertainty caused by sampling variance and instrument noise.

\section{Conclusions}
 In this study, we developed a complete simulation pipeline to assess systematic errors in measurements of the CMB by interferometers. Although we only focus on pointing errors at present, any other systematic errors, such as beam shape errors, gain errors and cross polarization, can be evaluated in the same way and we plan to study them in a forthcoming paper. The main purpose of the present paper is to introduce the maximum likelihood estimator for interferometric CMB temperature and polarization data, and to study its sensitivity to incorrect modeling assumptions and systematic errors. We choose tolerance levels of $10\%$ on $\alpha$ and $\beta$ in this study, which are somewhat arbitrary and may be changed at will. For interferometers with large numbers of redundant baselines, each independently measuring the Fourier modes of the sky, the effects of random systematic errors will be highly suppressed since errors will average down in visibilities measured over many baselines.

For a QUBIC-like interferometer, we find that in most cases the most stringent constraints on the allowable range of pointing errors are obtained from the requirement on the bias in power spectrum rather than from the changes in statistical errors. When the Gaussian-distributed pointing errors are controlled with a precision of $\delta\approx0.7^\circ$, our simulation shows that the measured $B$-modes in the multipole range $28<\ell<384$ can not be contaminated at the $10\%$ level in the statistical uncertainty ($\sigma_\ell$) units, but the change in the statistical error at $28<\ell<88$ could exceed $10\%$ of the statistical uncertainty. Our results are consistent with the analytical estimations by ~\cite{bunnsyst}, within a factor of 4.

As we know, the choice of scan strategy plays an important role in mitigating systematic effects. In imaging experiments, one uses natural sky rotation and frequent boresight rotation in order to achieve sufficient parallactic angle coverage and minimize systematic contamination in the B-mode power. For pointing errors, the recent study~\citep{shimon/etal:2008} shows that the leading order pointing effects would vanish for an ideal isotropic scan where every pixel is uniformly scanned in multiple random orientations. Unlike imaging experiments, interferometers rotate about the boresite in order to increase the $u$-$v$ coverage, provide a clean way to modulate the polarization signals and recover the Stokes parameters, and test for systematic effects using redundant baselines (e.g. DASI~\citep{kovac/etal:2002}). Other scan strategies such as continuous drift scans and mosaicking are valuable for reducing the sampling variance and improving the $\ell$-space resolution~\citep{White1999} while obtaining clean and optimal $E/B$ separation~\citep{Bunn:2007}. However, the pointing systematic effects for interferometers are insensitive to scanning strategies but sensitive to the configuration of the array elements for several reasons:  (1) interferometers measure the power spectrum directly and the underlying power spectrum is assumed to be the same in different sky patches; (2) any random systematics errors would be averaged out in visibilities measured by a large number of redundant baselines which only rely on the configuration of array elements; (3) the contamination only comes from a leakage of $E$ to $B$ (not $T$ to $B$);  (4) this leakage is independent of scanning strategies  -- that is, the errors do not strongly couple the visibilities at different angular scales to each other and the width of the coupling region for each Fourier mode is determined by the inverse of the beam width but not the scan strategy. Even if pointing errors do not have good statistical properties, the induced spurious polarization signals are still expected to be small if the total number of redundant baselines is large enough. For example, assuming the total $N(N-1)/2$ visibilities are measured by N antennas in which M antennas have pointing errors of $p=0.4$, according to Eq.~\ref{eq:tedan}, the resulting spurious $BB$ power would be $\propto p(M/N)$ and $\propto p(\sqrt{M}/N)$ in the fully-correlated and the uncorrelated case, respectively. Thus large pointing errors appearing only in a few antennas (e.g. $M \ll 400$ for QUBIC-like experiments) can not significantly bias the $BB$ measurements.

To evaluate how pointing errors limit the constraint on $r$, based on the simulated maps with input $BB$ of $r=0$, we compared the recovered $BB$ power spectra and the corresponding statistical errors in the lowest $\ell$-bin with and without pointing errors. We find that pointing errors with $p=0.1$ in both the fully-correlated and uncorrelated cases would slightly bias the 2-$\sigma$ upper limit on $r$ at $\sim 10\%$ level.

In principle the $TB$ and $EB$ power spectra are unique ``smoking gun'' signals for new physics. For imaging experiments, pointing errors have to be controlled to the sub-arcminute level to avoid spurious signals. However, our simulation clearly shows that the impact of the pointing errors on $TB$ and $EB$ estimates are negligibly small compared with their statistical uncertainties. Therefore systematic pointing errors in interferometers will not severely degrade B-mode science.


\section*{Acknowledgments}\label{con}
We acknowledge use of the FFTW library~\citep{fftw} and computing time and technical support from the University of Richmond under NSF Grant 0922748. LZ and PT acknowledge support from NSF Grant AST-0908900. GST and AK acknowledge support from NSF Grant AST-0908844. PMS acknowledges support from NSF Grant AST-0908902. BDW acknowledges funding from an ANR Chaire d'Excellence, the UPMC Chaire Internationale in Theoretical Cosmology, and NSF grants AST-0908902 and AST-0708849. EFB is supported by NSF Awards 0908319 and 0922748. We would like to thank an anonymous referee for valuable suggestions which helped us to significantly improve this paper.

\bibliography{max}
\bibliographystyle{apj}
\nocite{*}

\end{document}